\documentclass[aps]{revtex4}
\usepackage{eurosym}
\usepackage{amsfonts}
\usepackage{amsmath}
\usepackage{amssymb,epsf}
\usepackage{color}
\usepackage{graphicx}
\usepackage{epstopdf}
\usepackage{float}
\usepackage{caption}
\usepackage{subfig}

\begin{document}

\title{Two-dimensional Lifshitz-like AdS black holes in $F(R)$ gravity}
\author{B. Eslam Panah$^{1,2,3}$\footnote{
email address: eslampanah@umz.ac.ir} }
\affiliation{$^{1}$ Department of Theoretical Physics, Faculty of Science, University of
Mazandaran, P. O. Box 47416-95447, Babolsar, Iran\\
$^{2}$ ICRANet-Mazandaran, University of Mazandaran, P. O. Box 47416-95447,
Babolsar, Iran\\
$^{3}$ ICRANet, Piazza della Repubblica 10, I-65122 Pescara, Italy}

\begin{abstract}
Two-dimensional ($2D$) Lifshitz-like black holes in special $F(R)$ gravity
cases are extracted. We indicate an essential singularity at $r=0$, covered
with an event horizon. Then conserved and thermodynamic quantities such as
temperature, mass, entropy, and the heat capacity of $2D$ Lifshitz-like
black holes in $F(R)$ gravity are evaluated. Our analysis shows that $2D$
Lifshitz-like black hole solutions can be physical solutions, provided that
the cosmological constant is negative ($\Lambda<0$). Indeed, there is a
phase transition between stable and unstable cases by increasing the radius
of AdS black holes. In other words, the $2D$ Lifshitz-like AdS black holes
with large radii are physical and enjoy thermal stability. The obtained $2D$
Lifshitz-like AdS-black holes in $F(R)$ gravity turn to the well-known $2D$
Schwarzschild AdS-black holes when the Lifshitz-like parameter is zero ($s=0$%
). Also, correspondence between these black hole solutions and the $2D$
rotating black hole solutions is found by adjusting the Lifshitz-like
parameter.
\end{abstract}

\maketitle

\section{introduction}

One of the exciting observations in cosmology is related to the fact that
our Universe has an accelerated expansion \cite{expI,expIII}. Modified
theories of gravity are the simplest way to address this accelerating
behavior. In this regard, $F(R)=R+f(R)$ theory of gravity is the
straightforward modification of General Relativity (GR). The gravitational
action in this theory is a general function of $R$ \cite%
{F(R)I,F(R)II,F(R)III}. $F(R)$ gravity includes some exciting features such
as i) $F(R)$ models can be fixed according to the cosmological and
astrophysical observations ranging from local to cosmological scale (for
more details, see Refs. \cite{Mod1,Mod2,Mod4,Mod5,Mod5b,Mod6,Mod9}). ii) it
may explain the structure formation of the Universe without considering dark
energy or dark matter. iii) this theory of gravity is coincident with
Newtonian and post-Newtonian approximations \cite{CapozzielloI,CapozzielloII}%
. iv) $F(R)$ gravity may give the unified description of the early-time
inflation and late-time acceleration. v) the whole sequence of the
Universe's evolution epochs: inflation, radiation/matter dominance, and dark
energy may be extracted in this theory of gravity. According to the
mentioned features of $F(R)$ gravity, some people have studied different
solutions of this gravity in Refs. \cite%
{SoIII,SoV,SoVIII,SoIX,SoX,SoXI,SoXII}.

On the other hand, a new quantum theory of gravity based on Lifshitz's idea
in the quantum system was proposed by Horava \cite{HoravaI,HoravaII}, which
takes into account different spacetime footing. This theory is known as
Horava-Lifshitz gravity, which modifies GR in small-scale or ultraviolet
(UV) regimes. Notably, this theory reduces to GR in the infrared (IR) limit.
The Lorentz symmetry is broken based on this assumption. Consequently, the
theory is renormalizable at a quantum level. Indeed, Horava-Lifshitz gravity
introduces a new metric that includes an anisotropic scale invariant between
time and space as 
\begin{equation}
t\rightarrow b^{z}t~~~~~\&~~\ ~\ x\rightarrow bx,
\end{equation}%
where $z$ is a dynamical critical exponent that makes the theory
renormalizable for $z\geq d$ in $\left( d+1\right) -$dimensional spacetime.
However, Horava-Lifshitz's theory of gravity does not explain the dark
energy epoch in its original form as it also occurs in GR. To have a unified
theory of gravity in UV and IR regimes, $F(R)$ Horava-Lifshitz gravity has
been introduced. $F(R)$-Horava-Lifshitz's theory of gravity with different
points of view has been evaluated in Refs. \cite%
{FRHLI,FRHLII,FRHLIII,FRHLIV,FRHLVII,Kluson}.

Two-dimensional ($2D$) quantum theories of gravity were introduced to
understand the principles and puzzles of quantum gravity \cite%
{2II,2III,2IV,2V,2VI}. Such ideas of gravity are much simpler than
four-dimensional cases. Also, these theories share some exciting features
with four-dimensional gravity. Further, the $2D$ theories of gravity could
be identified with non-critical string theories, models with spacetime
super-symmetry have also been proposed in Refs. \cite{twoStI,twoStII}. In
this regard, $2D$ gravity with different motivations has been introduced,
and their properties are investigated in Refs. \cite%
{2D1,2D3,2D4,2D7,2D9,2D12,2D13,2D15,2D18,2D19,2D20}.

The main motivation for considering $F(R)$-Horava-Lifshitz's theory of
gravity is related to these facts that $F(R)$ gravity modifies GR in the
infrared (IR) limit, and also Horava-Lifshitz gravity modifies GR in
small-scale or ultraviolet (UV) regime. On the other hand, to understand
quantum gravity, $2D$ theories of gravity can be helpful. So the
investigation of $2D$ black hole solutions in the context of such a unified
theory of gravity can give us some essential information. Therefore we
consider $2D$ Lifshitz-like spacetime and extract black hole solutions in $%
F(R)$ gravity. Notably, the black hole solutions in four \cite{Lif4D}, and
three \cite{HendiR} dimensional Lifshitz-like spacetime have been studied in 
$F(R)$ gravity.

The structure of this paper is as follows: at first, we introduce the
fundamental equation of $F(R)$ gravity. Then we obtain $2D$ Lifshitz-like
black holes in the context of $F(R)$ gravity. In the next section, we study
thermodynamic properties and the heat capacity of these black holes. Then,
we find that the $2D$ Lifshitz-like AdS-black holes in $F(R)$ gravity may
recover the Schwarzschild black holes and the $2D$ rotating-like black holes
in $2D$ spacetime by adjusting the Lifshitz-like parameter. The final
section is devoted to concluding remarks.

\section{Basic Equations}

The action of $F(R)$ gravity in $2D$ spacetime is given by \cite{NojiriOI} 
\begin{equation}
I=\frac{1}{2\kappa^{2}}\int d^{2}x\sqrt{-g}F(R),  \label{action}
\end{equation}%
where $\kappa^{2}=8\pi G$, and $G$ is the Newtonian gravitational constant.
Also, $g=det(g_{\mu\nu})$ is the determinant of metric tensor $g_{\mu \nu }$%
. Here, we set the Newtonian gravitational and the speed of light equal to $%
1 $ ($G=c=1$). The above action explains a theory of $2D$ gravity where $%
F(R)=R+f(R)$ and $f(R)$ is a arbitrarily function of the Ricci scalar $R$.
It is worthwhile to mention that for $F(R)=R$, the action (\ref{action})
recovers the Hilbert-Einstein action.

The field equations are obtained by variation with respect to metric tensor $%
g_{\mu \nu }$, in the following form 
\begin{equation}
R_{\mu \nu }F_{R}-\nabla _{\mu }\nabla _{\nu }F_{R}+\left( \Box F_{R}-\frac{1%
}{2}F(R)\right) g_{\mu \nu }=\mathrm{0},  \label{FE}
\end{equation}%
where $F_{R}\equiv dF(R)/dR$, and ${R}_{\mu \nu }$ is the Ricci tensor.
Also, $\Box=\nabla^{\nu }\nabla _{\nu }$.

\section{$2D$ Lifshitz-like black holes}

We consider the following $2D$ Lifshitz spacetime ansatz%
\begin{equation}
ds^{2}=-\left( \frac{r^{2}}{l^{2}}\right) ^{z}h\left( r\right) dt^{2}+\frac{%
l^{2}dr^{2}}{r^{2}h\left( r\right) },  \label{Metric}
\end{equation}%
where $z$ is a real number called the Lifshitz parameter, and $l$ is an
arbitrary positive length scale. To extract exact solutions, we apply the
following transformations, 
\begin{equation}
\left( \frac{r^{2}}{l^{2}}\right) h(r)\rightarrow g\left( r\right)
,~~~~~\&~~~~~l\rightarrow r_{0},~~~~\ \&~~2z-2\rightarrow s,  \notag
\end{equation}%
in which the metric (\ref{Metric}) becomes 
\begin{equation}
ds^{2}=-\left( \frac{r}{r_{0}}\right) ^{s}g\left( r\right) dt^{2}+\frac{%
dr^{2}}{g\left( r\right) },  \label{Metric2}
\end{equation}%
where $s$ and $r_{0}$ are Lifshitz-like parameter and an arbitrary positive
length scale, respectively. In order to obtain exact $2D$ Lifshitz-like
black hole solutions in $F(R)$ gravity, we consider the metric (\ref{Metric2}%
), and the field equation (\ref{FE}). In addition, to have a theory
renormalizable in two-dimensional spacetime, we must respect to $s\geq 0$
(or $z\geq 1$).

It is worthwhile to mention that in this paper, we intend to extract $2D$
black hole solutions in Lifshitz-like spacetime for the special class of $%
F(R)$ gravity which their Ricci scalars are constant, i.e., $R=R_{0}=$
constant. Indeed, we use the mentioned method in Ref. \cite{Calza}, which is
applicable for $F(R)$ models with two constraints, simultaneously, $%
F(R_{0})=0$ and $F_{R}=0$. This class of $F(R)$ models does not lead to the
usual GR since the vacuum field equation of GR identically satisfies $F_{R}=1
$ with vanishing Ricci scalar. Some of the important features of
this model of $F(R)$ gravity are

1) The early-time inflation or late-time accelerated expansion fall
in this class of theories. In other words, this theory of gravity can unveil
some cosmological scenarios that cannot be described in the frame of
standard general relativity \cite{Calza}.

2) This class of $\mathbf{F(R)}$ theories of gravity may
explain the anomalous rotation curve typical of spiral galaxies. These
models can also mimic the influence of the dark matter halo that is supposed
to surround spiral galaxies. In addition, in this case there is a potential
for fixing the parameters of the theory by analyzing the rotation curves of
a large number of galaxy \cite{Calza}. 

3) There are a few solutions in this theory that lead to traversable
wormholes. The conceptual advantage of these solutions is the fact that
there is no need for unnatural matter sources that violate general
relativity energy conditions \cite{Calza}.

4) The charged (A)dS black hole solutions were extracted without
considering the matter field and the cosmological constant. Indeed, the
electrical charge and the cosmological constant create as a result of the
geometry of spacetime in this theory \cite{SoVIII}.

5) We can find some new solutions to the equations of motion of this
class of $F(R)$ theories of gravity without solving them
explicitly. In the case of black holes, the metric is often different from
the standard Schwarzschild one. These new solutions lead to interesting
results \cite{HendiR,Bertipagani2021}.

Using the metric (\ref{Metric2}), we can obtain the metric function for $%
R=R_{0}$, from 
\begin{equation}
R_{0}=g{^{\prime \prime }+}\frac{3sg{^{\prime }}}{2r}+\frac{s\left(
s-2\right) g}{2r^{2}},  \label{R0}
\end{equation}%
where $g=g(r)$, $g{^{\prime }=}\frac{dg(r)}{dr}$, and $g{^{\prime \prime }=}%
\frac{d^{2}g(r)}{dr^{2}}$. Considering the Eq. (\ref{R0}) and after some
calculations, we can find an exact solution in the following form 
\begin{equation}
g(r)=-\frac{m}{r^{\frac{s}{2}-1}}+\frac{c}{r^{s}}-\Lambda r^{2},
\label{totalg}
\end{equation}%
where $m$ and $c$ are two integration constants. Also, $\Lambda $ is a
constant which depends on $R_{0}$, as $\Lambda =\frac{2R_{0}}{\left(
s+2\right) ^{2}}$. In addition, the Kretschmann scalar can be written in the
following form 
\begin{equation}
\mathcal{K}=R_{\alpha \beta \mu \nu }R^{\alpha \beta \mu \nu }=\frac{\left(
s+2\right) ^{2}\Lambda ^{2}}{4}.
\end{equation}

In order to find a singularity of the obtained solution (\ref{totalg}), we
calculate all of the components of Ricci and Riemannian tensors. The Ricci
tensors are 
\begin{equation}
R_{~t}^{t}=R_{~r}^{r}=\frac{\mathcal{K}}{\Lambda },~~~~\&~~~~R^{tt}=\frac{%
R_{tt}}{g^{2}(r)\left( \frac{r}{r_{0}}\right) ^{2s}},~~~~%
\&~~~~R^{rr}=g^{2}(r)R_{rr},
\end{equation}%
where $R_{tt}=-\frac{\mathcal{K}g(r)}{\Lambda }\left( \frac{r}{r_{0}}\right)
^{s}$ and $R_{rr}=\frac{\mathcal{K}}{\Lambda g(r)}$. It is evident that $%
R_{rr}$ diverges\ at $r=0$ (i. e., $\underset{r\rightarrow 0}{\lim }%
R_{rr}\rightarrow \infty $), by considering $s=0$ and $s>0$. Besides for the
Riemannian tensor, we have 
\begin{equation}
\left\{ 
\begin{array}{ccc}
\underset{r\rightarrow 0}{\lim }R_{~rtr}^{t}=\frac{\mathcal{K}}{\Lambda g(r)}%
\rightarrow \infty &  & s=0 \\ 
&  &  \\ 
\underset{r\rightarrow 0}{\lim }R^{trtr}=-\frac{\mathcal{K}}{\Lambda }\left( 
\frac{r}{r_{0}}\right) ^{-s}\rightarrow \infty &  & s>0%
\end{array}%
\right. ,
\end{equation}%
where confirms that there is at least a singularity at $r=0$ for different
values of $s$. In addition, according to the Eq. (\ref{totalg}), the
asymptotical behavior of the obtained solution is Anti-de Sitter (AdS) or de
Sitter (dS) spacetime for $\Lambda <0$ or $\Lambda >0$, respectively.

In order to find roots of the solution (\ref{totalg}), we solve $g(r)=0$,
which becomes 
\begin{equation}
r_{\pm }=\left( \frac{2\Lambda }{\left[ -m\mp \sqrt{m^{2}+4\Lambda c}\right] 
}\right) ^{-2/\left( 2+s\right) },  \label{root}
\end{equation}%
which $r_{-}$ and $r_{+}$ are the inner and outer (event) horizons,
respectively. To have real and positive event horizon ($r_{+}$), we should
respect to $m^{2}>-4\Lambda c$, and $\Lambda <0$. As a result, the $2D$
Lifshitz-like black holes have real and positive event horizons when $%
\Lambda <0$. In other words, according to our finding, the $2D$
Lifshitz-like AdS-black holes can be valid in $F(R)$ gravity. As a result,
the solution (\ref{totalg}) includes a singularity at $r=0$, which is
covered by an event horizon. In other words, the solution (\ref{totalg}) is
related to the $2D$ AdS-black hole in Lifshitz-like spacetime.

To more investigate the obtained $2D$ Lifshitz-like AdS-black holes in $F(R)$
gravity, we determine the admissible space of the parameters of the metric
function (\ref{totalg}). For this purpose, we can do it by studying the
extremal black hole criteria 
\begin{equation}
g(r_{+})=0=g{^{\prime }}\left( r_{+}\right) .  \label{extermal}
\end{equation}

The equation (\ref{extermal}) indicates that the metric function has one
degenerate horizon at $r_{+}$, which corresponds to the coincidence of the
inner and outer horizons. Solving these two equations simultaneously, leads
to 
\begin{equation}
m=\frac{2c}{r_{+}^{\frac{s}{2}+1}},~~~~~\&~~~~~\Lambda =\frac{-c}{r_{+}^{s+2}%
}.  \label{m}
\end{equation}

\begin{figure}[tbh]
\centering
\includegraphics[width=0.38\linewidth]{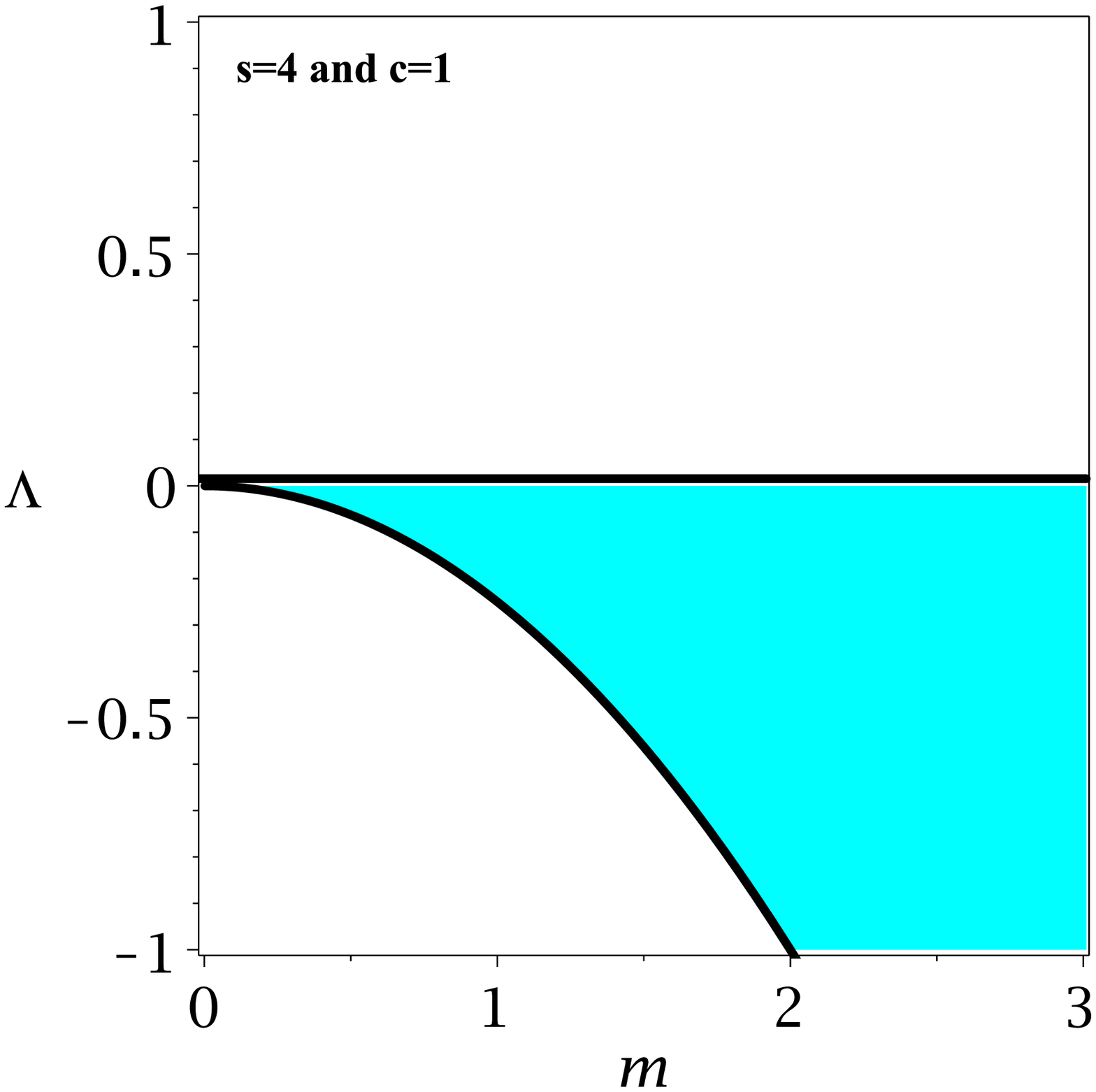} \includegraphics[width=0.4%
\linewidth]{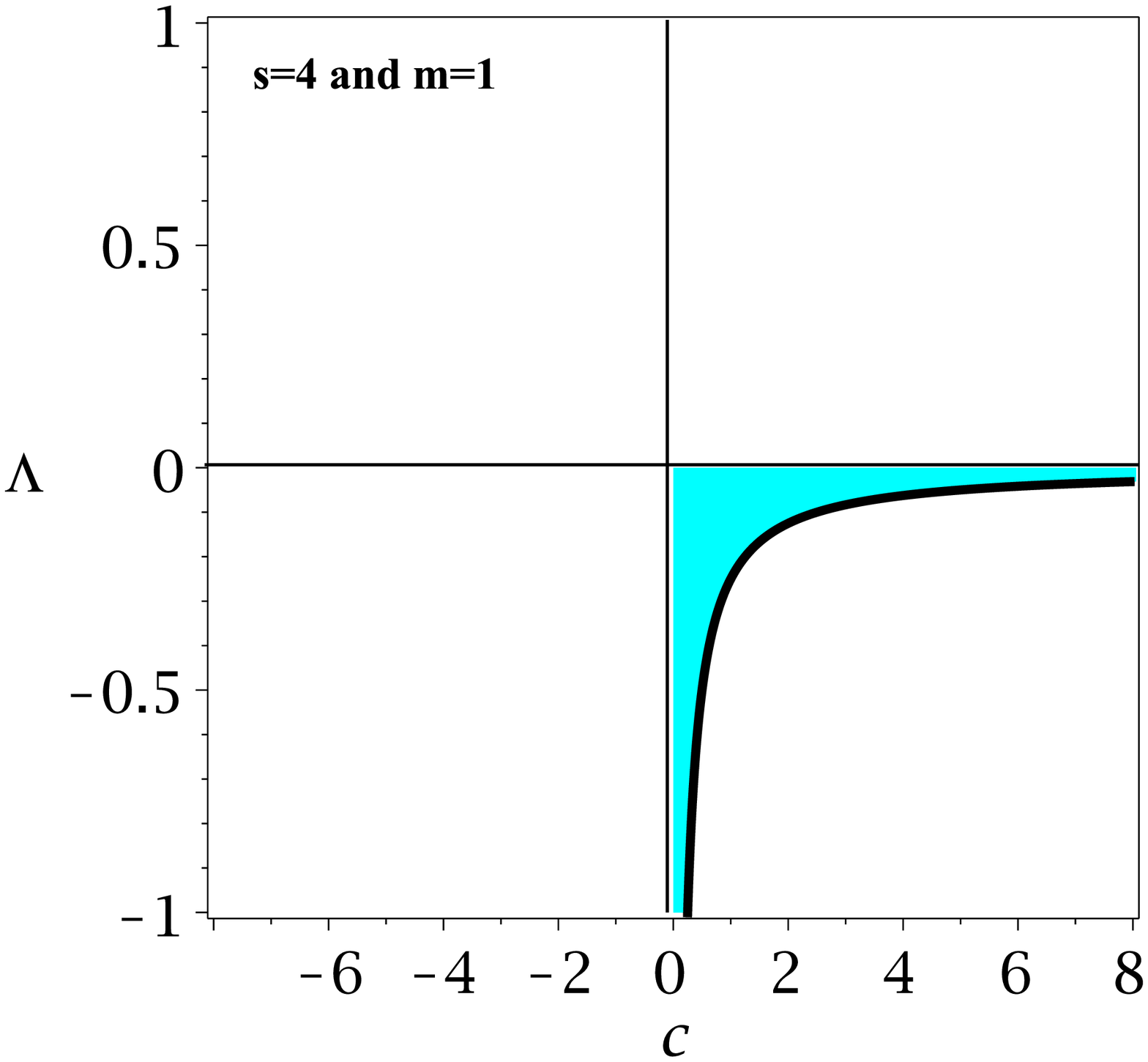}
\caption{The admissible parameter space of metric function ($g(r)$).}
\label{Fig1}
\end{figure}

Using the equation (\ref{m}), we can plot the admissible parameter space of
metric function in the $\Lambda -m$ plan. The resultant curve is depicted in
the left panel of Fig. \ref{Fig1}, by the black line and the blue area. In
the left panel of Fig. \ref{Fig1}, the black line and the blue area denote
the extremal limit and black holes with two horizons (inner and event
horizons), respectively. Notably, no black hole exists out of the black line
and the blue area. Our results show that there are only black holes with the
negative values of the cosmological constant ($\Lambda <0$). In the same
way, by using the equation (\ref{extermal}), one can determine the
admissible parameter space of metric function in the $\Lambda -c$ plan (see
the right panel in Fig. \ref{Fig1}). We use the following equations for the $%
\Lambda -c$ plan 
\begin{equation}
c=\frac{m}{2}r_{+}^{\frac{s}{2}+1},~~\ ~~\&~~~~~\Lambda =\frac{-m}{2r_{+}^{%
\frac{s}{2}+1}}.
\end{equation}

It is clear that the $2D$ Lifshitz-like black holes only exist for the
negative value of the cosmological constant ($\Lambda <0$).

%%%%%%%%%%%%%%%%%%%%%%%%%%%%%%%%%%%%%%%%%%%%%%%%%%%%%%%%%%%%%%%
%%%%%%%%%%%%%%%%%%%%%%%%%%%%%%%%%%%%%%%%%%%%%%%%%%%%%%%%%%%%%%%

\begin{figure}[tbh]
\centering
\includegraphics[width=0.4\linewidth]{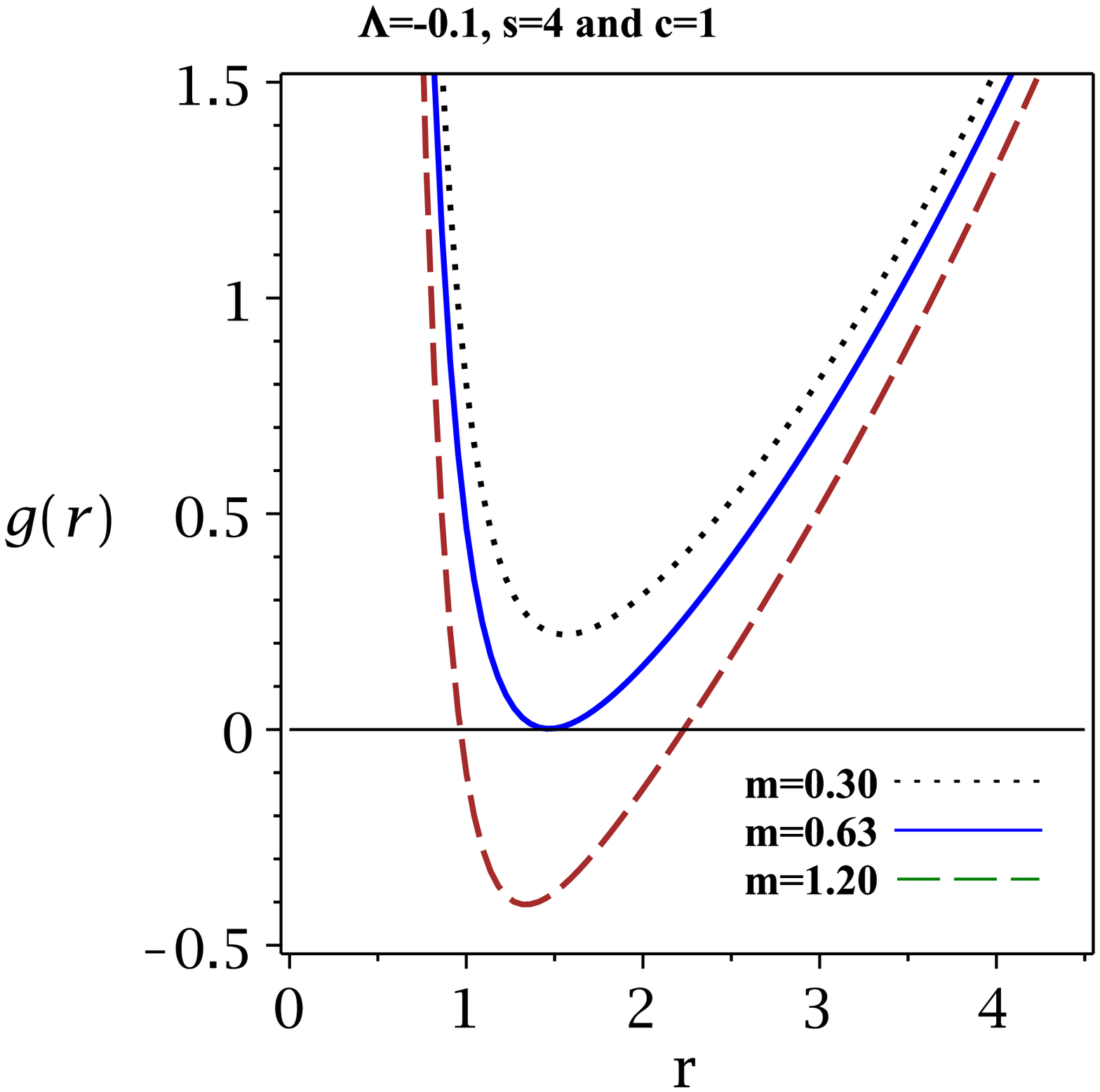} \includegraphics[width=0.4%
\linewidth]{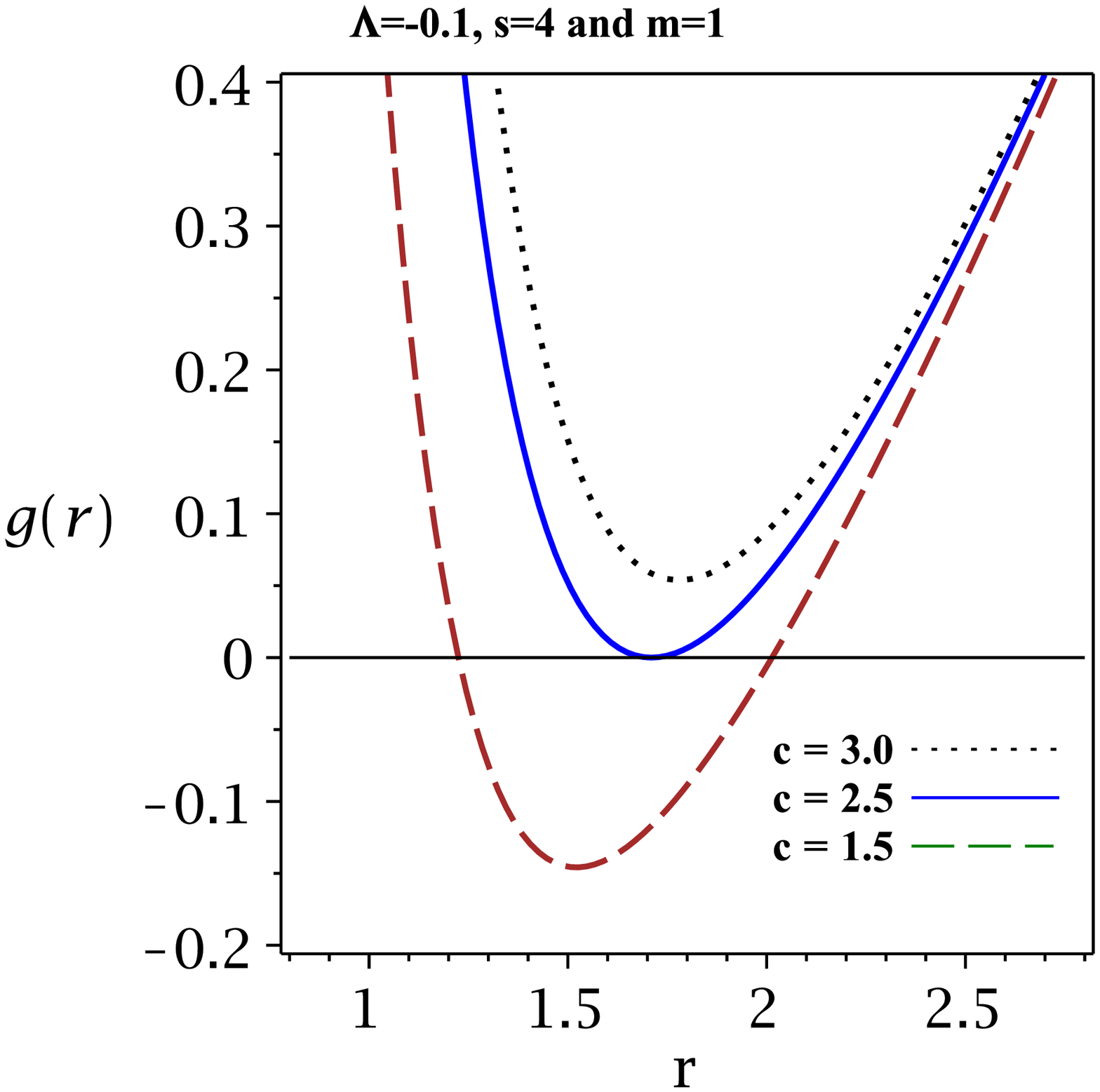} \includegraphics[width=0.4\linewidth]{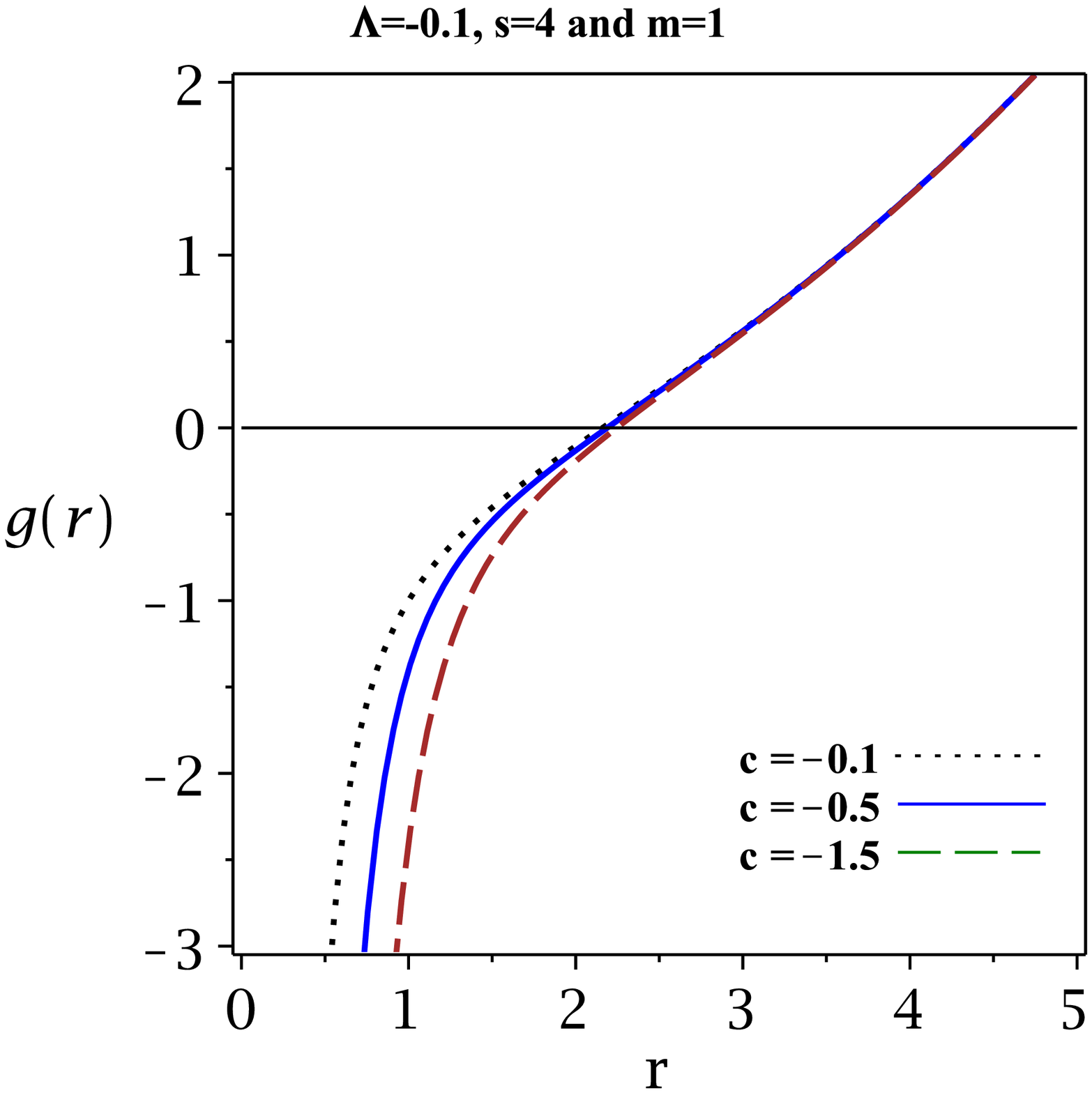} %
\includegraphics[width=0.4\linewidth]{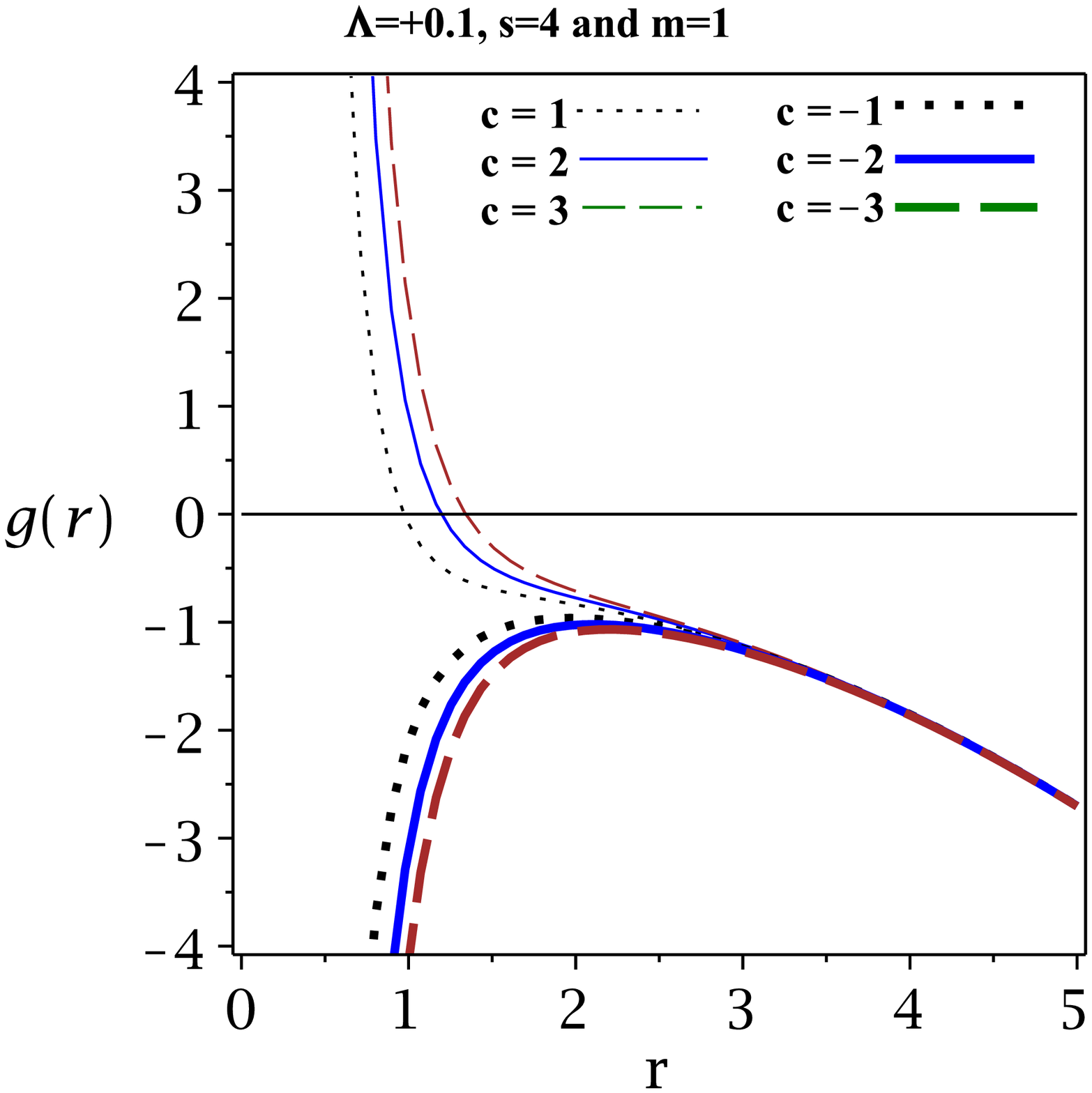}
\caption{$g(r)$ versus $r$ for different values of $m$, $\Lambda $, and $%
c_{1}$.}
\label{Fig2}
\end{figure}

To more investigate the behavior of the obtained solution (\ref{totalg}), we
plot the metric function versus $r$ in Fig. \ref{Fig2}. Similar to
Reissner-Nordstr\"{o}m black holes, by adjusting the values of free
parameters, the metric function (\ref{totalg}) may have two roots (black
hole with an inner horizon and an outer (event) horizon), an extreme root
(extremal black hole) or no root (naked singularity: no black hole). Indeed,
for $\Lambda <0$ and $c>0$, the black hole may have two roots (see the up
panels in Fig. \ref{Fig2}). For $\Lambda <0$ and $c<0$, there is one root
(see the down left panel in Fig. \ref{Fig2}). On the other hand, for $%
\Lambda >0$ with $c>0$ (or $c<0$), there is no event horizon. Also, by
considering $r_{+}=r_{-}$ in Eq. (\ref{root}), the $2D$ Lifshitz-like
extreme black hole is obtained as 
\begin{equation}
m=\pm 2\sqrt{-\Lambda c},
\end{equation}%
where the above relation reveals that for the $2D$ Lifshitz-like extreme
black holes in AdS spacetime, $c$ has to be positive.

In short, our analysis determines that black holes in the $2D$ Lifshitz-like
spacetime are AdS, and also the extreme case exists when $c>0$. 

\section{Conserved and thermodynamic quantities and thermal stability}

In this section, we study the thermodynamic properties of the $2D$
Lifshitz-like AdS-black holes. To calculate thermodynamic quantities, we
start with the Hawking temperature. The Hawking temperature of the black
hole on the event horizon ($r_{+}$) may be obtained through the use of the
definition of surface gravity, 
\begin{equation}
T=\frac{1}{2\pi }\sqrt{\frac{-1}{2}\left( \nabla _{\mu }\chi _{\nu }\right)
\left( \nabla ^{\mu }\chi ^{\nu }\right) },
\end{equation}%
where $\chi $ is the Killing vector. It is notable that the mentioned
spacetime contains a temporal Killing vector ($\chi =\partial /\partial t$).
Now we can obtain the Hawking temperature of the $2D$ Lifshitz-like
AdS-black hole in the following form 
\begin{equation}
T=\frac{-\left( s+2\right) }{8\pi }\left( r_{+}\Lambda +\frac{c}{r_{+}^{s+1}}%
\right) \left( \frac{r_{+}}{r_{0}}\right) ^{s/2}.  \label{T}
\end{equation}

After some calculation, we can find the root of temperature (\ref{T}), as 
\begin{equation}
r_{\text{root(}T\text{)}}=\left( \frac{-\Lambda }{c}\right) ^{-1/(s+2)}.
\label{RootT}
\end{equation}

\begin{figure}[tbh]
\centering
\includegraphics[width=0.4\linewidth]{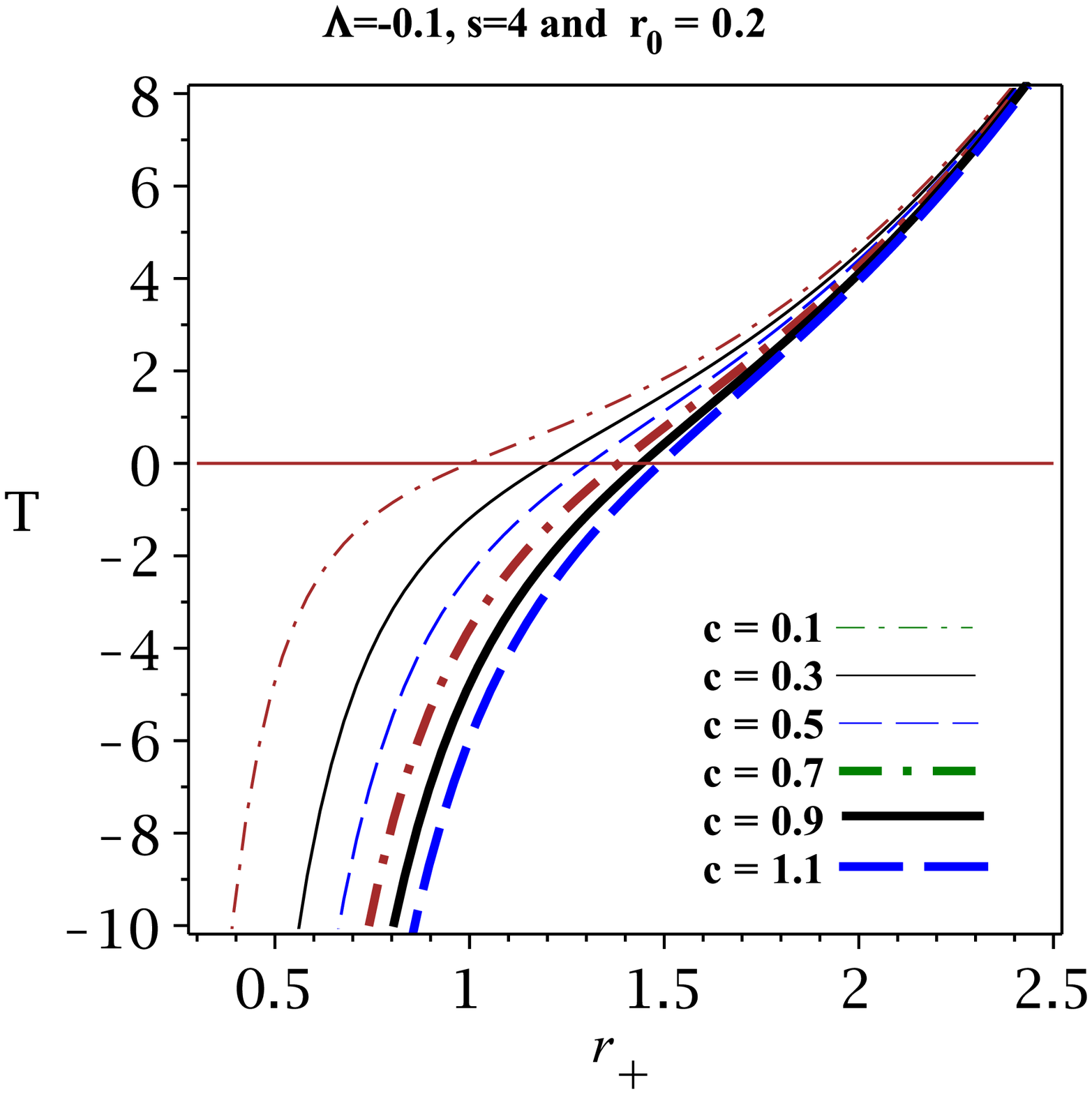} \includegraphics[width=0.4%
\linewidth]{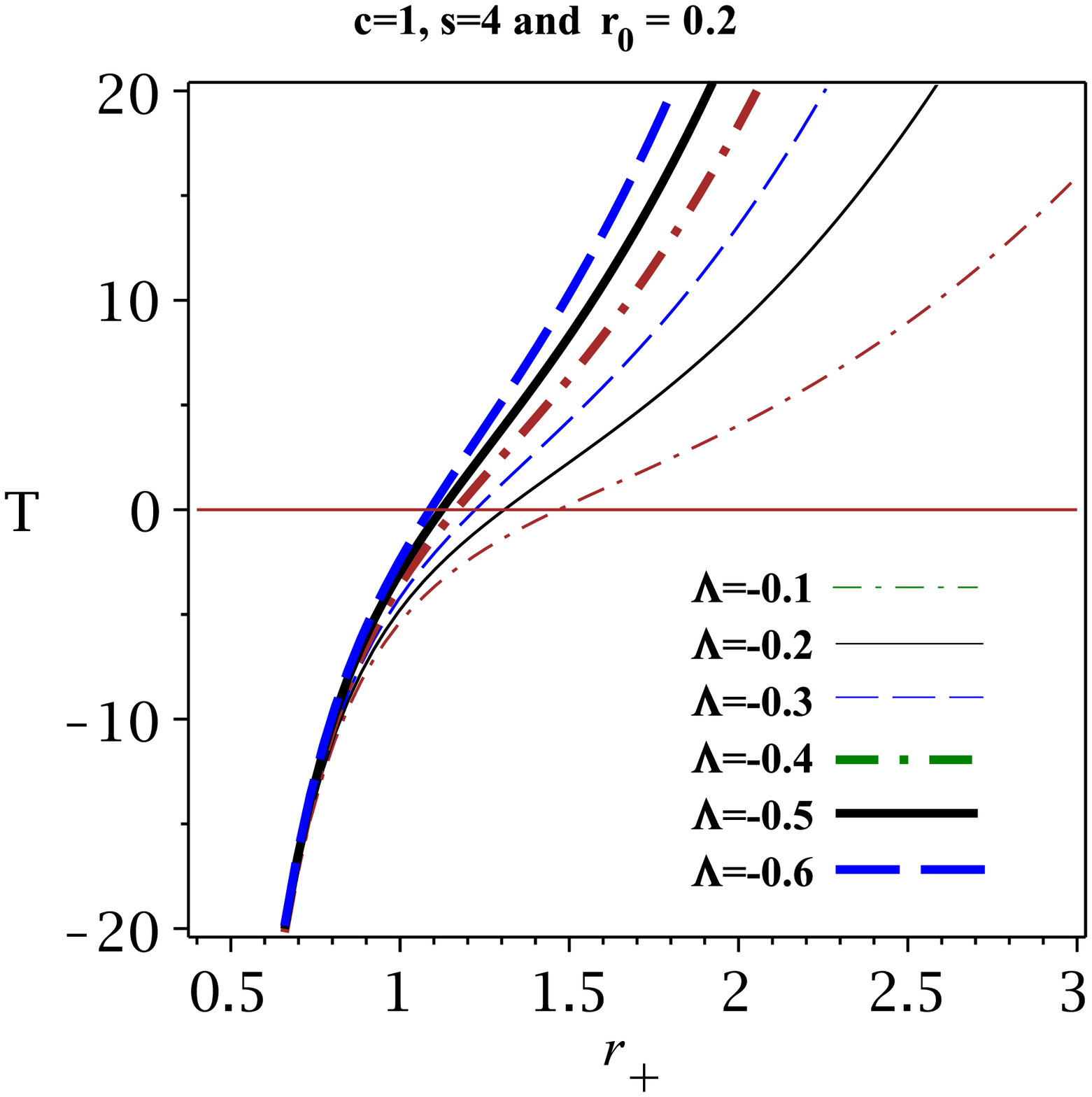} \includegraphics[width=0.4\linewidth]{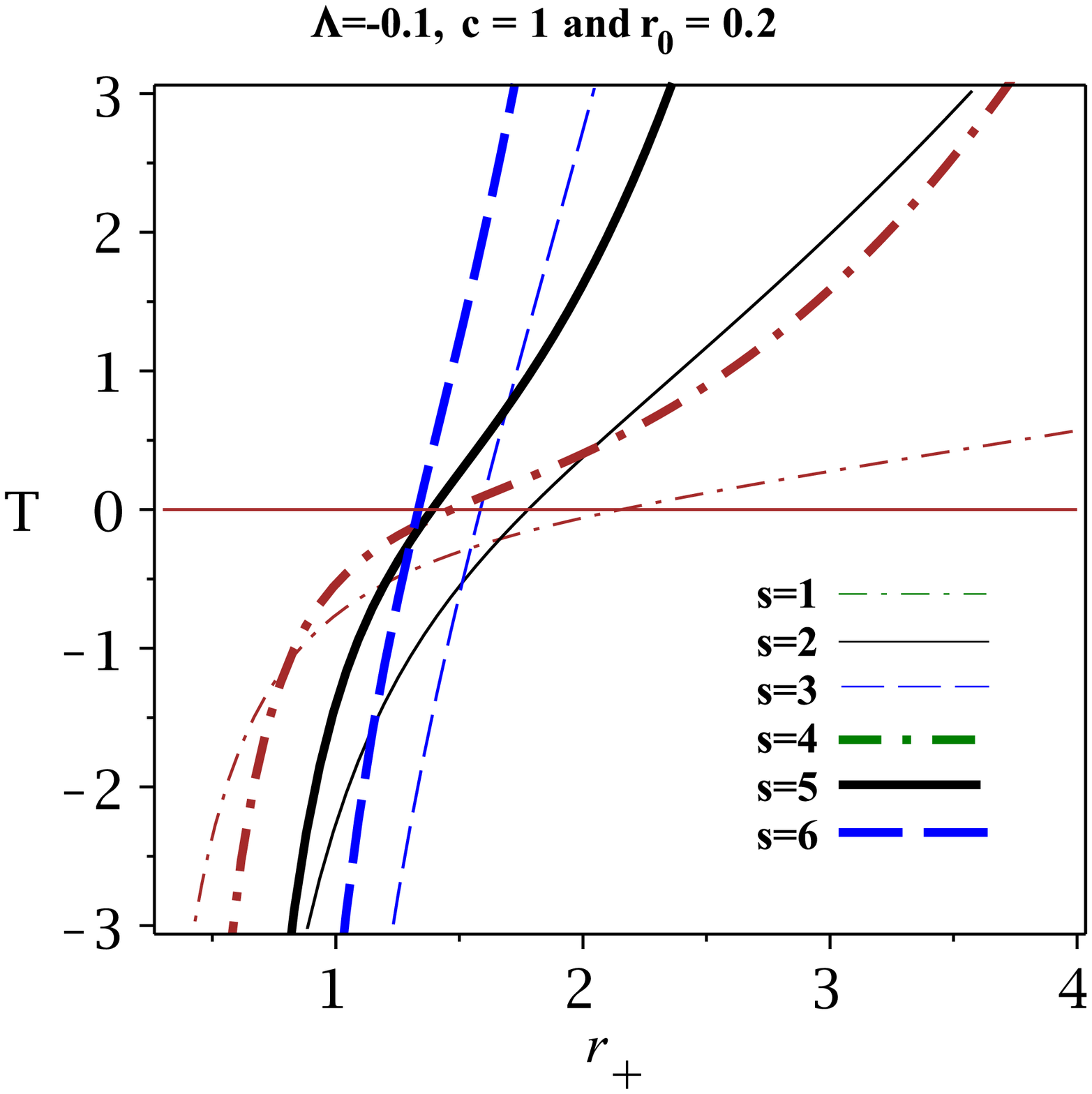} %
\includegraphics[width=0.4\linewidth]{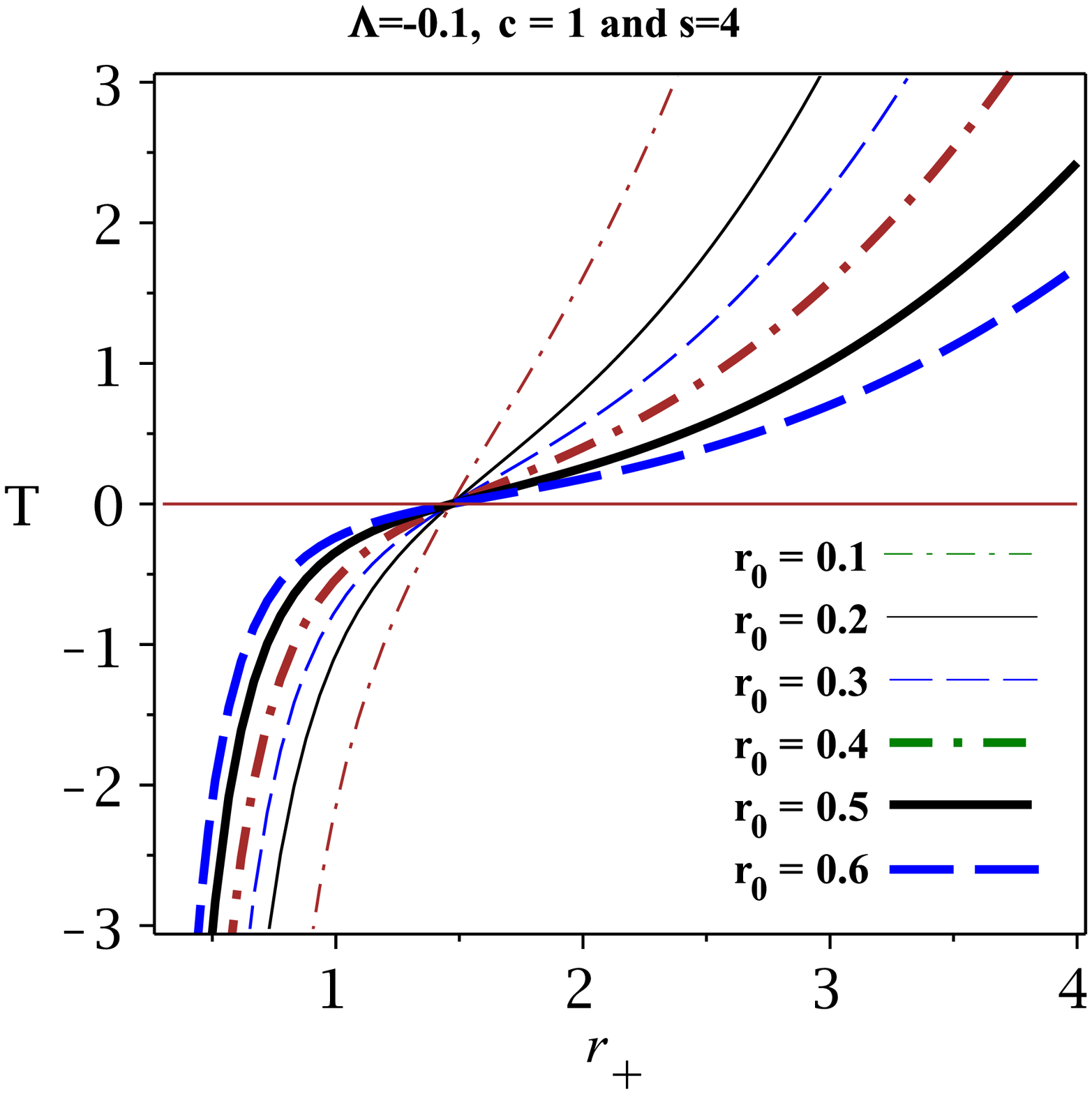}
\caption{$T$ versus $r_{+}$ for different values of $c$ (up left panel), $%
\Lambda$ (up right panel), $s$ (down left panel) and $r_{0}$ (down
rightpanel).}
\label{Fig3}
\end{figure}

According to the equation (\ref{RootT}), in order to have a real and
positive root, we have to the following conditions $\Lambda <0$, and $c>0$.
We plot the temperature (\ref{T}) versus $r_{+}$ in Fig. \ref{Fig3}.

%%%%%%%%%%%%%%%%%%%%%%%%%%%%%%%%%%%%%%%%%%%%%%%%%%%%%%%%%%%%%%%
%%%%%%%%%%%%%%%%%%%%%%%%%%%%%%%%%%%%%%%%%%%%%%%%%%%%%%%%%%%%%%%

The equation (\ref{RootT}) and Fig. \ref{Fig3}, show that the temperature
root ($r_{\text{root(}T\text{)}}$) of the $2D$ Lifshitz-like AdS-black holes
becomes large by increasing (decreasing) $c$ and $\Lambda $ ($s$), see the
up panels and down left panel in Fig. \ref{Fig3}. But for different value of 
$r_{0}$, the root of temperature dose not change (see the down right panel
in Fig. \ref{Fig3}).

To obtain the finite mass of $2D$ Lifshitz-like AdS-black holes, we use the
Ashtekar-Magnon-Das (AMD) formula for a far away observer \cite{AMDI,AMDII}.
Considering the AMD approach, the total mass can be written as \cite%
{AMDI,AMDII,AMDIII,AMDIV} 
\begin{equation}
M=m_{0}=mr_{0}^{1-\frac{s}{2}},
\end{equation}%
where $m$ is related to the geometrical mass ($m=m_{0}r_{0}^{\frac{s}{2}-1}$%
) of the metric function (\ref{totalg}). Using the above relation and the
metric function (\ref{totalg}) on the event horizon ($g\left( r_{+}\right)=0 
$), one can obtain $M$ as a function of $r_{+}$ as 
\begin{equation}
M=\left( \frac{c}{r_{+}^{s}}-\Lambda r_{+}^{2}\right) \left( \frac{r_{+}}{%
r_{0}}\right) ^{\frac{s}{2}-1}.  \label{M}
\end{equation}

\begin{figure*}[tbh]
\centering
\includegraphics[width=0.32\linewidth]{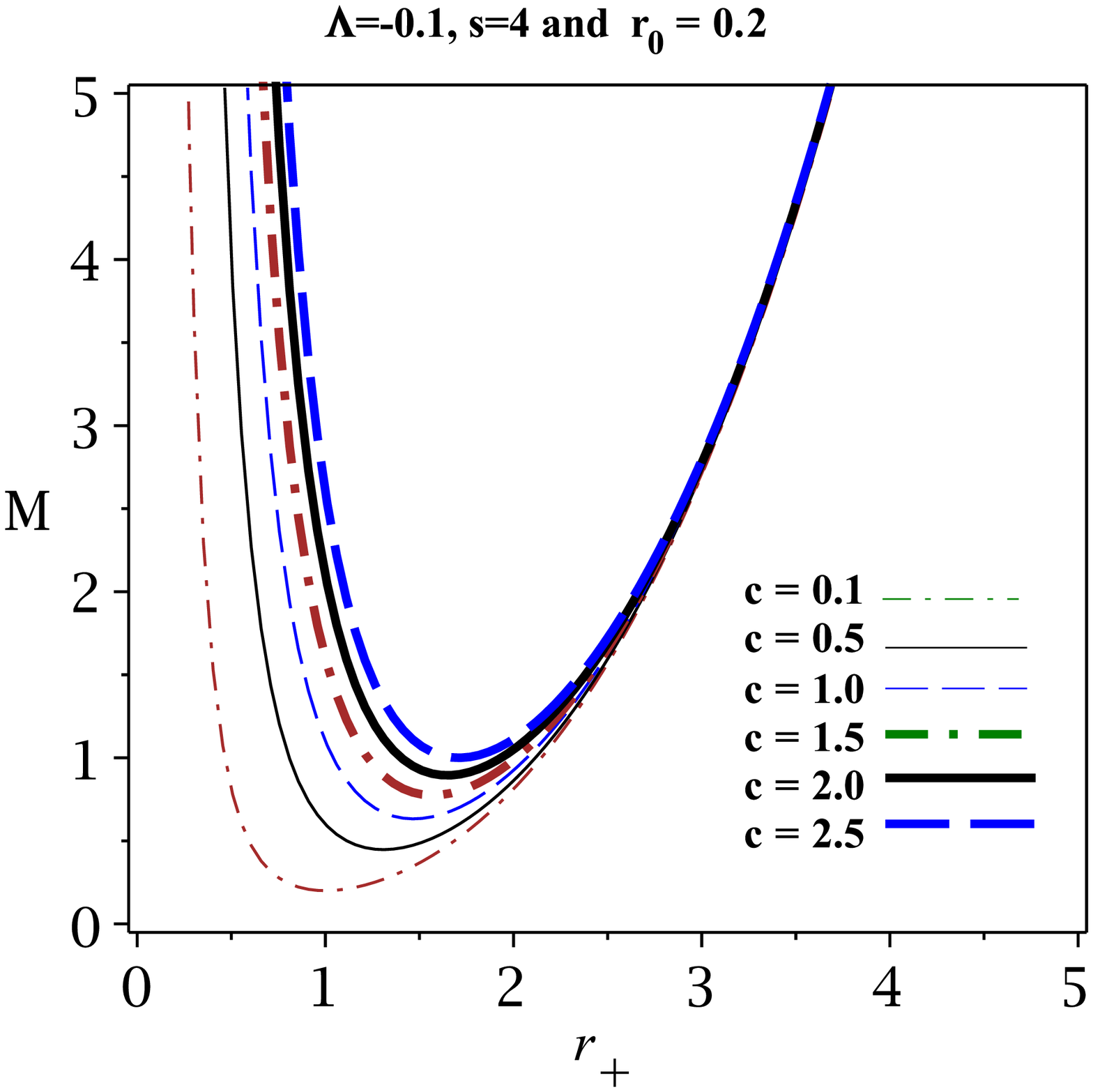} \includegraphics[width=0.32%
\linewidth]{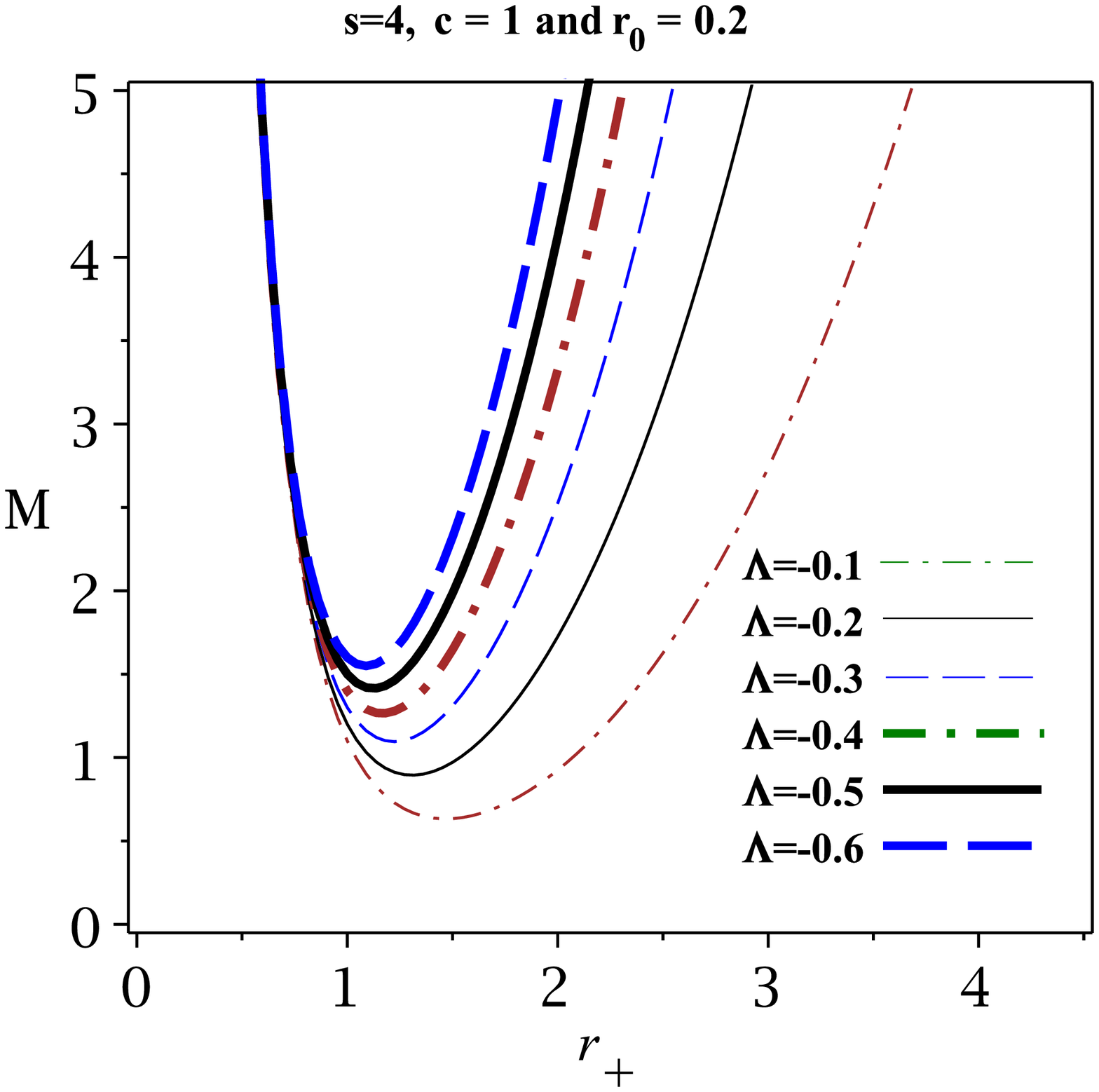} \includegraphics[width=0.32\linewidth]{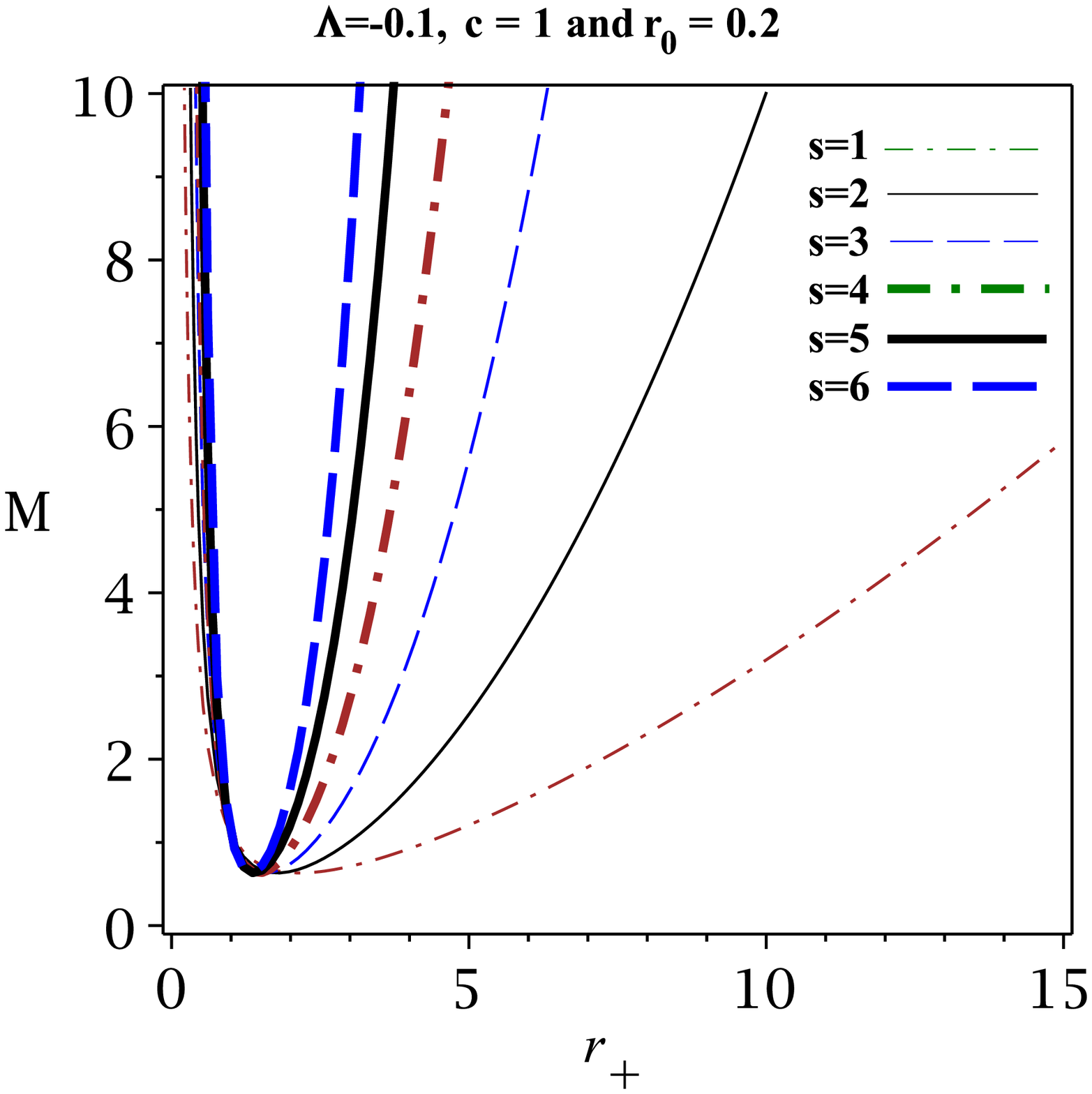}
\caption{$M$ versus $r_{+}$ for different values of $c$ (left panel), $%
\Lambda$ (middle panel) and $s$ (right panel).}
\label{Fig4}
\end{figure*}

Considering the obtained mass (\ref{M}), we have plotted the mass of the $2D$
Lifshitz-like AdS-black holes versus $r_{+}$, in Fig. \ref{Fig4}.

%%%%%%%%%%%%%%%%%%%%%%%%%%%%%%%%%%%%%%%%%%%%%%%%%%%%%%%%%%%%%%%
%%%%%%%%%%%%%%%%%%%%%%%%%%%%%%%%%%%%%%%%%%%%%%%%%%%%%%%%%%%%%%%

The results in Fig. \ref{Fig4}, show that the mass of $2D$ Lifshitz-like
AdS-black hole is positive. Also, there is a minimum mass for this black
hole, where changes for different values of $c$ and $\Lambda $ (see the left
and middle panels in Fig. \ref{Fig4}). But this minimum does not change for
different values of $s$ (see the right panel in Fig. \ref{Fig4}).

To extract the entropy of the $2D$ Lifshitz-like AdS-black holes, one can
respect the validity of thermodynamics first law. Therefore we can obtain
the entropy as 
\begin{equation}
\delta S=\frac{1}{T}\delta M,
\end{equation}%
by inserting Eqs. (\ref{T}) and (\ref{M}), in the above equation, the
entropy is obtained as 
\begin{equation}
S=\int \frac{1}{T}\delta M=\frac{4\pi r_{0}}{s+2}\ln \left( \frac{r_{+}}{%
r_{0}}\right) ^{s+2}=4\pi r_{0}\ln \left( \frac{r_{+}}{r_{0}}\right) ,
\label{S}
\end{equation}%
where is independent of the Lifshitz-like parameter ($s$). It is notable
that the entropy in $F(R)$ gravity is extracted by using generalized area
law as $S=\frac{A}{4}F_{R}=\frac{A}{4}\left( 1+f_{R}\right) $ in Ref. \cite%
{Entropy}, where $A$ is the event horizon area of the black holes. But here,
we have used two constraints $F(R)=0$ and $F_{R}=0$, and therefore, the
generalized area law leads to zero entropy. So, to obtain the entropy, we
considered the thermodynamics first law.

The heat capacity is one of the thermodynamical quantities carrying
important information regarding the black holes thermal structure in the
canonical ensemble. Notably, the discontinuities of this quantity mark the
possible thermal phase transitions that the system can undergo. Also, the
heat capacity's (negativity) positivity determines whether the system is
thermally (in)stable. In addition, the roots of this quantity may give
information about the possible changes between stable/unstable states or
bound points \cite{HeatI}. Due to these critical points, we evaluate the $2D$
Lifshitz-like AdS-black holes' heat capacity to study the thermal structure
of such black holes. Indeed, we will indicate that by using the heat
capacity alongside the temperature, we can draw a picture regarding
stability/instability of the $2D$ Lifshitz-like AdS-black holes in $F(R)$
gravity.

Using the Eqs. (\ref{T}) and (\ref{S}), one can obtain the heat capacity in
the following form 
\begin{equation}
C=T\left( \frac{\partial S}{\partial T}\right) =T\frac{\left( \frac{\partial
S}{\partial r_{+}}\right) }{\left( \frac{\partial T}{\partial r_{+}}\right) }%
=\frac{8\pi \left( \Lambda r_{+}^{s+2}+c\right) r_{0}}{\left( \Lambda
r_{+}^{s+2}-c\right) \left( s+2\right) },
\end{equation}%
which has two interesting points, namely the root ($r_{\text{root}}$) and
divergence ($r_{\text{div}}$) points, where are 
\begin{equation}
r_{\text{root}}=\left( \frac{-\Lambda }{c}\right) ^{-1/(s+2)},~~~~~\&~~~~\
r_{\text{div}}=\left( \frac{\Lambda }{c}\right) ^{-1/\left( s+2\right) }.
\label{rootdiv}
\end{equation}

\begin{figure}[tbh]
\centering
\includegraphics[width=0.4\linewidth]{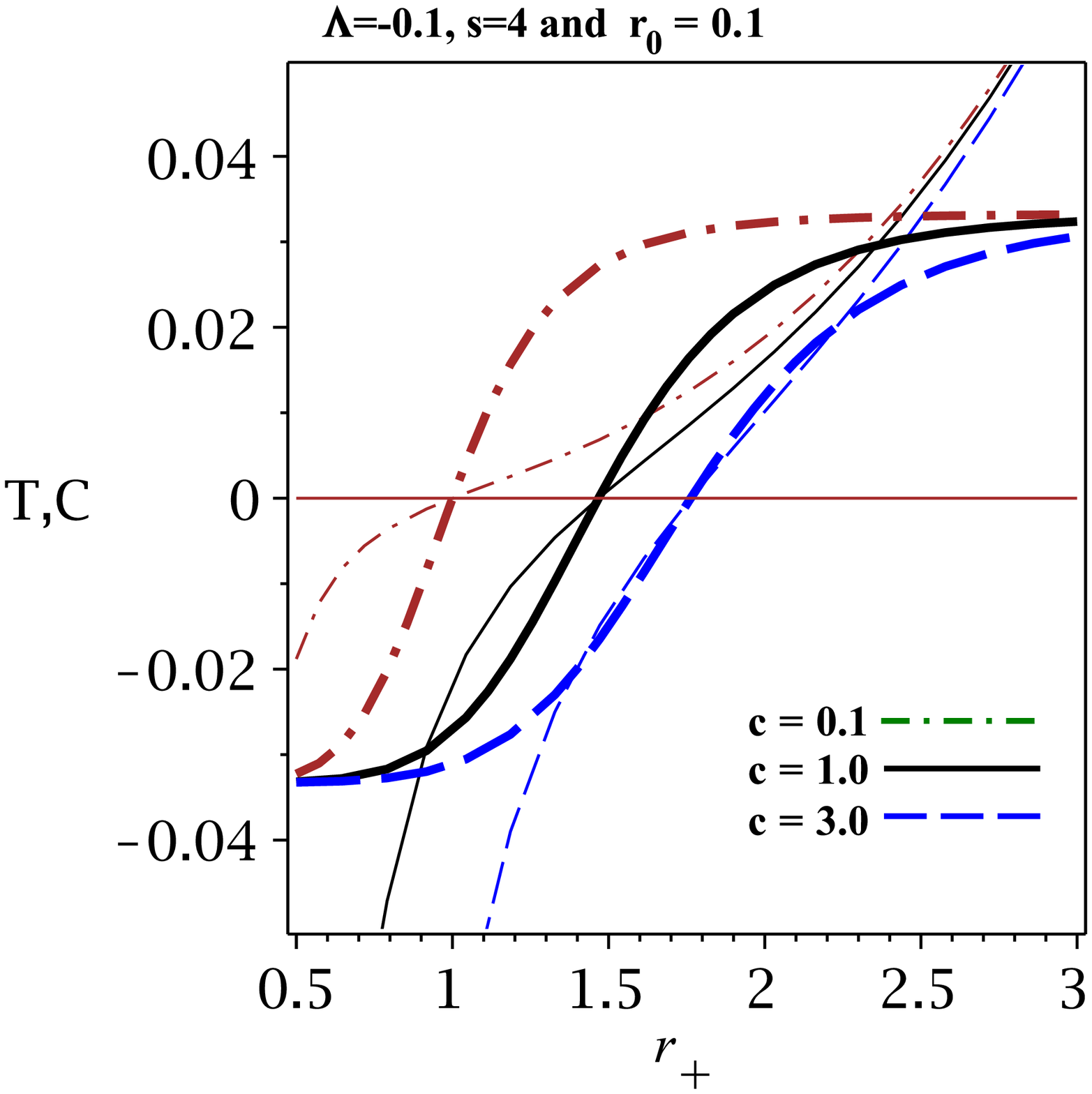} \includegraphics[width=0.4%
\linewidth]{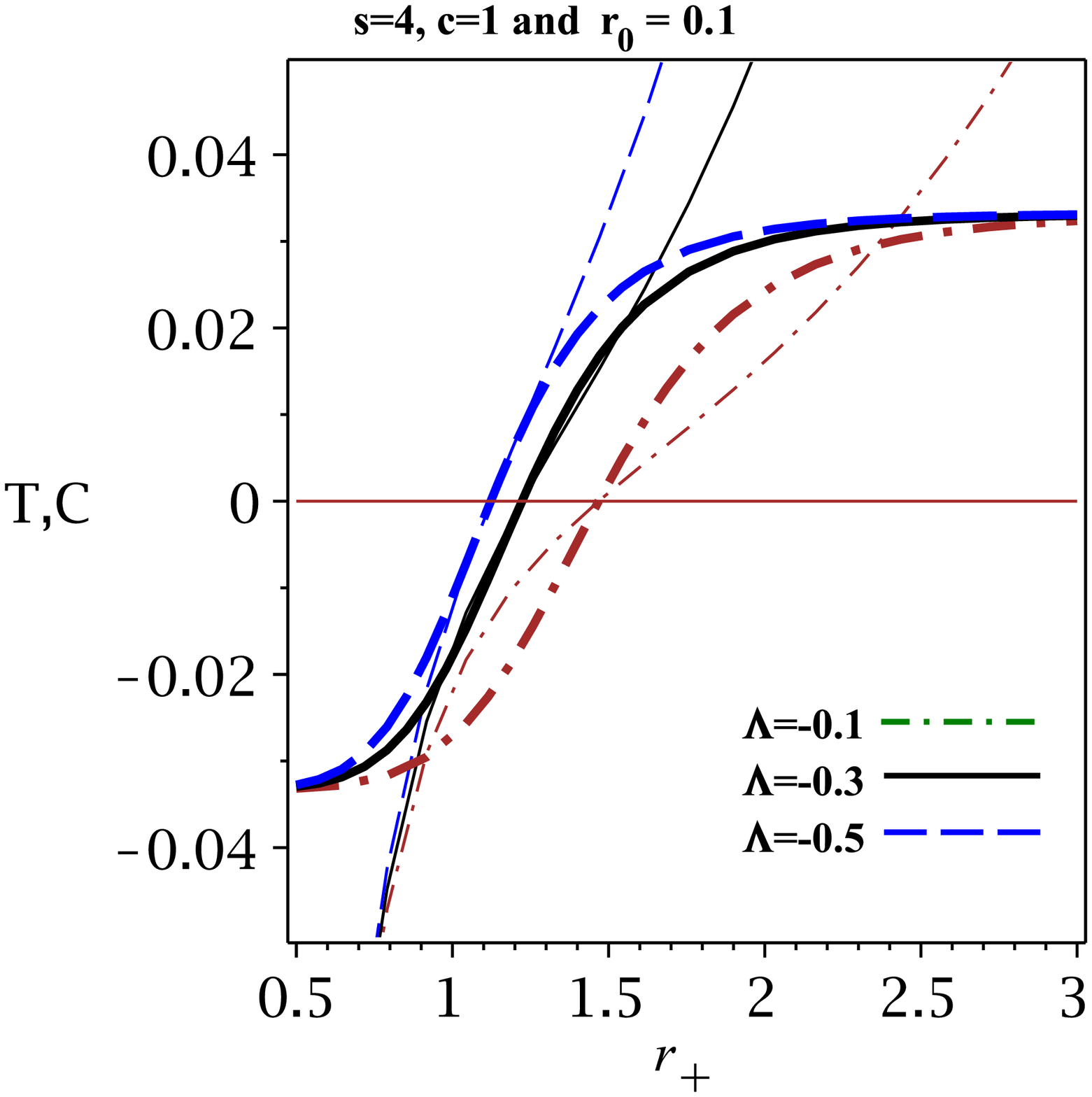} \includegraphics[width=0.4\linewidth]{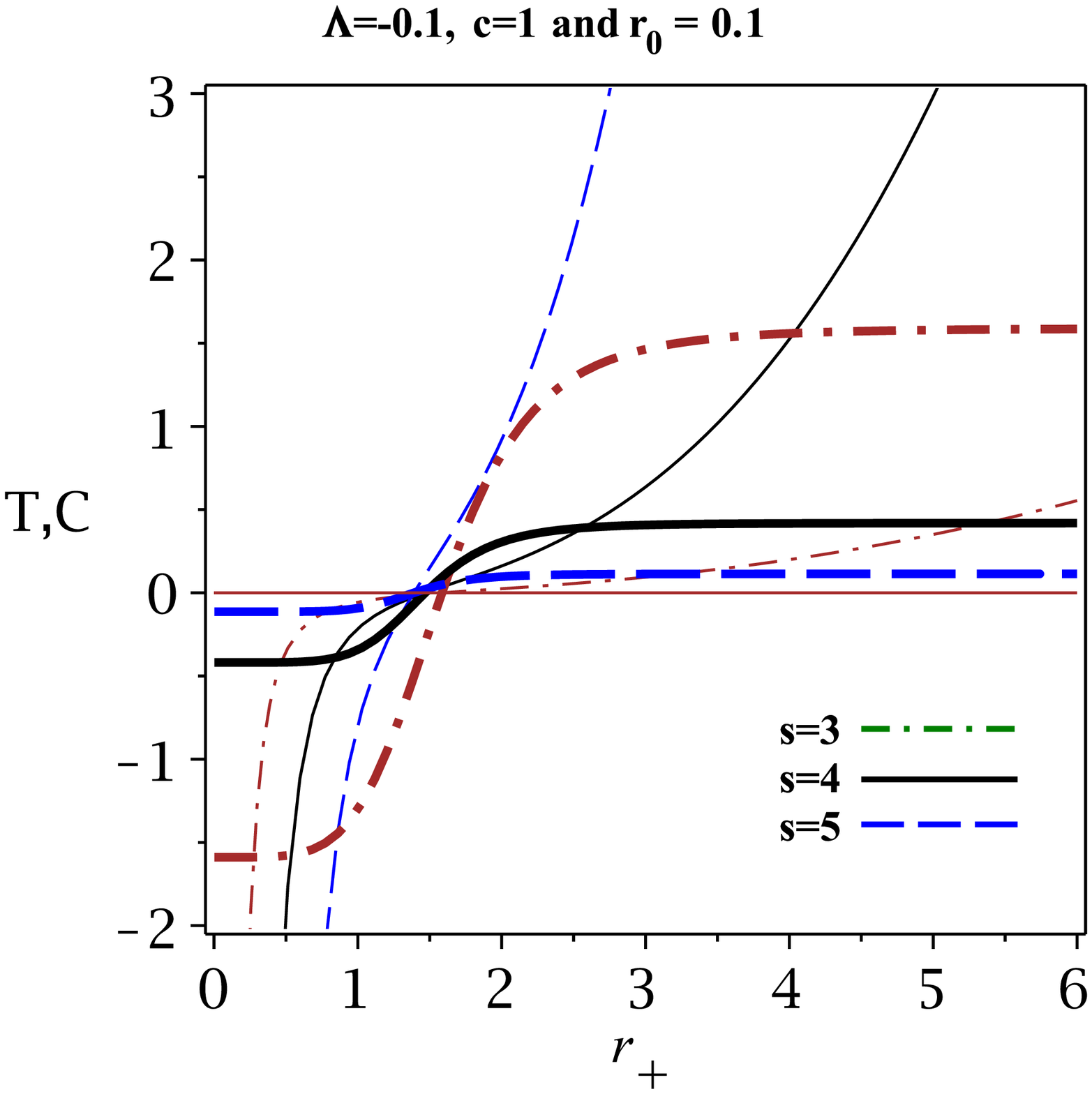} %
\includegraphics[width=0.4\linewidth]{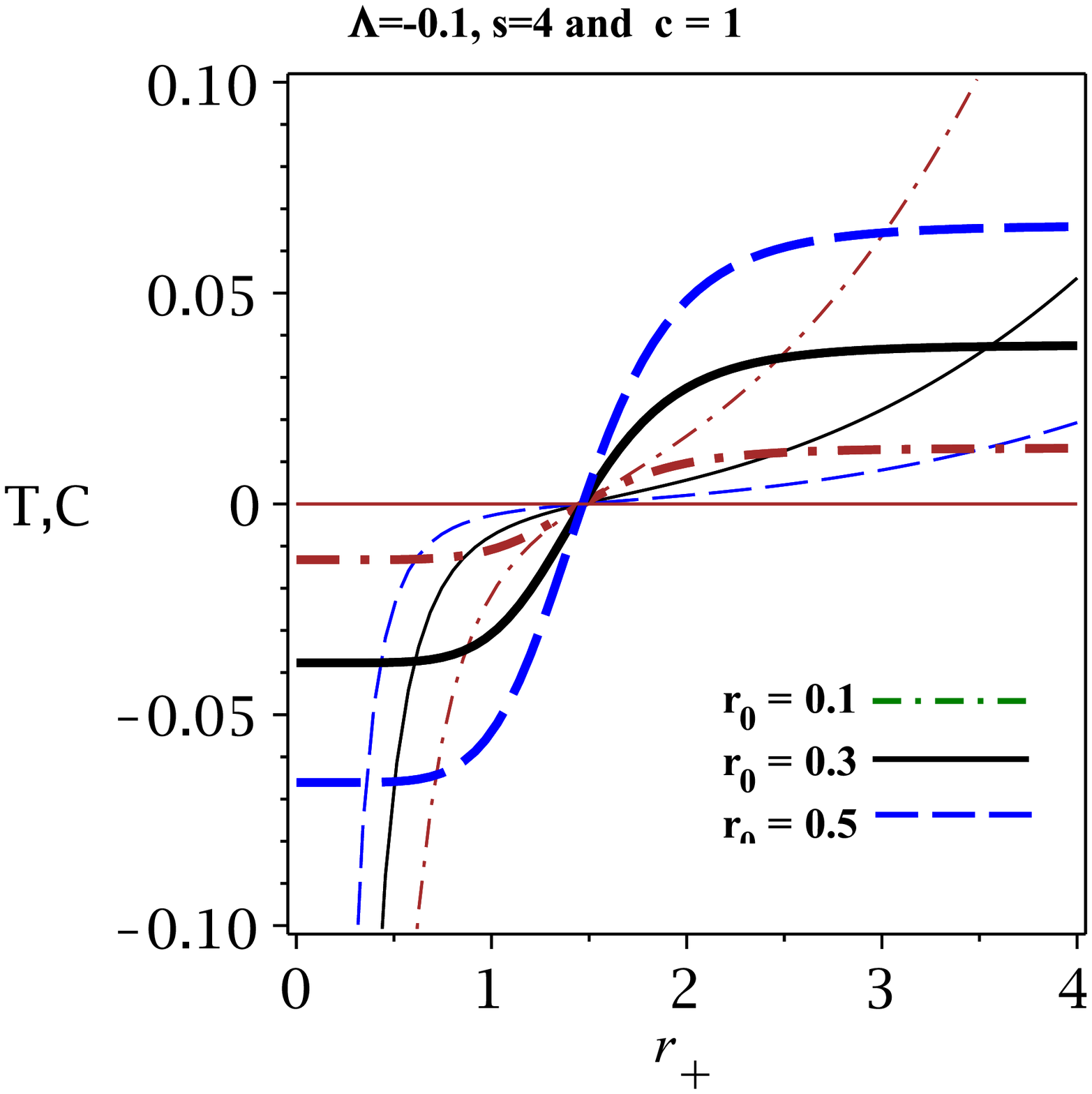}
\caption{$C$ (thick lines) and $T$ (thin lines) versus $r_{+}$ for different
values of $c$ (up left panel), $\Lambda$ (up right panel), $s$ (down left
panel) and $r_{0}$ (down right panel).}
\label{Fig5}
\end{figure}

According to Eq. (\ref{rootdiv}), there is not any divergence point ($r_{%
\text{div}}$) for the $2D$ Lifshitz-like AdS-black holes, since $\Lambda$ is
negative. To more study the behavior of heat capacity, we plot it in Fig. %
\ref{Fig5}. Our solutions indicate that exists one root ($r_{\text{root}}$)
for the temperature and the heat capacity. Indeed, there is one critical
root, namely $r_{\text{root}}$, in which for $r_{+}<r_{\text{root}}$, the
temperature and heat capacity of the $2D$ Lifshitz-like black holes are
negative. Therefore, the black holes with small radii are non-physical and
unstable. Nevertheless, for $r_{+}>r_{\text{root}}$, the temperature and the
heat capacity of these black holes are positive. In other words, the $2D$
Lifshitz-like AdS-black holes with large radii (i.e., $r_{+}>r_{\text{root}}$%
) are physical and enjoy thermal stability. In addition, the physical and
thermal stability areas of these black holes decrease by increasing $c$ (up
left panel in Fig. \ref{Fig5}) and $\Lambda $ (up right panel in Fig. \ref%
{Fig5}), and also by increasing $s$, this area increases (down left panel in
Fig. \ref{Fig5}). On the other hand, different values of $r_{0}$, do not
affect this area (down right panel in Fig. \ref{Fig5}).

%%%%%%%%%%%%%%%%%%%%%%%%%%%%%%%%%%%%%%%%%%%%%%%%%%%%%%%%%%%%%%%
%%%%%%%%%%%%%%%%%%%%%%%%%%%%%%%%%%%%%%%%%%%%%%%%%%%%%%%%%%%%%%%

As a result, the obtained $2D$ Lifshitz-like AdS-black holes in $F(R)$
gravity with large radii are physical and enjoy thermal stability.

In the following, different values of Lifshitz-like parameters are evaluated
for the $2D$ Lifshitz-like AdS-black holes in $F(R)$ gravity. Indeed, we
indicate that the obtained $2D$ Lifshitz-like AdS-black hole in $F(R)$
gravity turns to the well-known $2D$ Schwarzschild AdS-black hole when the
Lifshitz-like parameter is zero, $s=0$. Besides, there is a correspondence
between this black hole solution and the $2D$ rotating black hole solution
when the Lifshitz-like parameter is two, $s=2$.

\subsection{Case $s=0$: $2D$ Schwarzschild black holes}

The $2D$ Schwarzschild AdS-black holes are obtained by adjusting $s=0$ in
the metric function (\ref{totalg}). In other words, by replacing $s=0$ in
Eq. (\ref{totalg}), we have 
\begin{equation}
g_{\underset{s=0}{}}(r)=-mr+c-\Lambda r^{2},  \label{caseI}
\end{equation}%
where the solution (\ref{caseI}) is known as $2D$ black holes of the
Schwarzschild-AdS spacetime when $c=1$ \cite{NojiriOI}. So, we have 
\begin{equation}
g_{\underset{s=0}{}}(r)=-mr+1-\Lambda r^{2}.  \label{caseIa}
\end{equation}

Considering the obtained solution (Eq. (\ref{caseIa})), we look for the
essential singularity(ies). The Ricci and the Kretschmann scalars and also $%
tt$-component of the Ricci tensor are given 
\begin{equation}
R_{0}=2\Lambda ,~~~~~\&~~~~~\mathcal{K}=\Lambda ^{2},~~~~~\&~~~~~R_{tt}=%
\frac{-\Lambda }{mr-1+\Lambda r^{2}},
\end{equation}%
which $\underset{r\rightarrow 0}{\lim }R_{tt}\rightarrow \infty $. So, there
is a curvature singularity at $r=0$.

Two roots of the solution (\ref{caseIa}), are given by $g(r)=0$, as 
\begin{equation}
r\pm =\frac{-m\mp \sqrt{m^{2}+4\Lambda }}{2\Lambda }.
\end{equation}

\begin{figure}[tbh]
\centering
\includegraphics[width=0.4\linewidth]{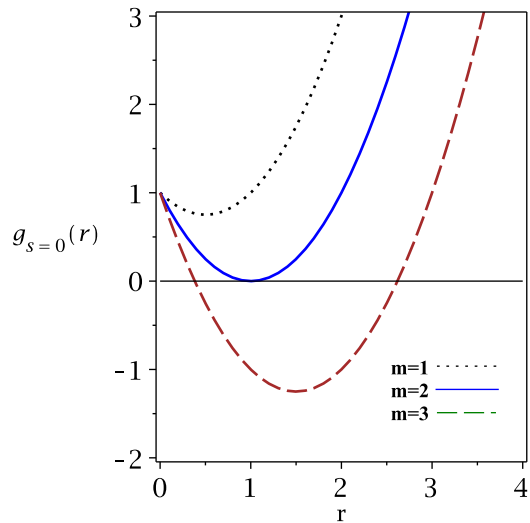}
\caption{$g_{\protect\underset{s=0}{}}(r)$ versus $r$ for $\Lambda =-1$, and
different values of $m$.}
\label{Fig6}
\end{figure}

Another interesting result is the existing two roots for the $2D$
Schwarzschild AdS-black holes. As we know, there is one root for
Schwarzschild black holes in four-dimensional spacetime, whereas the $2D$
Schwarzschild-like AdS black holes have two roots (similar to
Reissner-Nordstr\"{o}m black holes). In order to study the $2D$
Schwarzschild-like AdS-black holes in more details, we plot $g_{\underset{s=0%
}{}}(r)$ versus $r$ in Fig. \ref{Fig6}.

%%%%%%%%%%%%%%%%%%%%%%%%%%%%%%%%%%%%%%%%%%%%%%%%%%%%%%%%%%%%%%%
%%%%%%%%%%%%%%%%%%%%%%%%%%%%%%%%%%%%%%%%%%%%%%%%%%%%%%%%%%%%%%%

Figure. \ref{Fig6}, shows that by increasing the parameter $m$, the black
holes have two horizons. In other words, massive $2D$ Schwarzschild-like
AdS-black holes may have two horizons.

We calculate the Hawking temperature, mass, entropy, and the heat capacity
of the $2D$ Schwarzschild-like AdS-black holes in the following forms 
\begin{eqnarray}
T_{_{s=0}} &=&-\frac{1+\Lambda r_{+}^{2}}{4\pi r_{+}},~~~~~%
\&~~~~~M_{_{s=0}}=\left( 1-\Lambda r_{+}^{2}\right) \left( \frac{r_{0}}{r_{+}%
}\right) ,  \notag \\
&& \\
S_{_{s=0}} &=&4\pi r_{0}\ln \left( \frac{r_{+}}{r_{0}}\right)
,~~~~~\&~~~~~C_{_{s=0}}=\frac{4\pi \left( \Lambda r_{+}^{2}+1\right) r_{0}}{%
\Lambda r_{+}^{2}-1}.  \notag
\end{eqnarray}

\begin{figure}[tbh]
\centering
\includegraphics[width=0.4\linewidth]{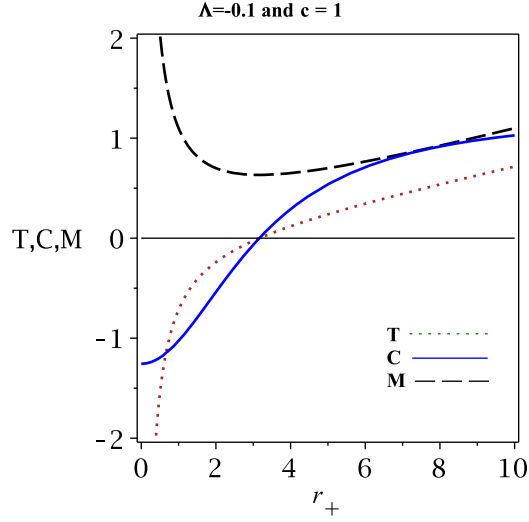}
\caption{$T$ (dotted line), $M$ (dashed line), $C$ (continuous line) versus $%
r_{+}$.}
\label{Fig7}
\end{figure}

To study the behaviors of temperature, mass, and heat capacity, we plot Fig. %
\ref{Fig7}.

%%%%%%%%%%%%%%%%%%%%%%%%%%%%%%%%%%%%%%%%%%%%%%%%%%%%%%%%%%%%%%%
%%%%%%%%%%%%%%%%%%%%%%%%%%%%%%%%%%%%%%%%%%%%%%%%%%%%%%%%%%%%%%%

Our findings in Fig. \ref{Fig7}, indicate that the $2D$ Schwarzschild-like
AdS-black holes with large radii are physical and enjoy thermal stability.
In other words, for $r_{+}>r_{\text{root}}$, the temperature and the heat
capacity are positive. But for $r_{+}<r_{\text{root}}$, the temperature and
the heat capacity of these black holes are negative.

\subsection{Case $s=2$: $2D$ rotating-like black hole}

By replacing $s=2$, in the presented solution (\ref{totalg}), we have 
\begin{equation}
g_{\underset{s=2}{}}(r)=-m-\Lambda r^{2}+\frac{c}{r^{2}}.  \label{caseII}
\end{equation}

It is worthwhile to mention that there is an interesting correspondence
between the black hole solution (\ref{caseII}) and the $2D$ rotating black
hole (which is obtained from an appropriate, effective action for looking at 
$S-$wave scattering off a spinning BTZ black hole \cite{Achucarro}) when $c=%
\frac{J^{2}}{4}$. So, by considering $c=\frac{J^{2}}{4}$, the solution (\ref%
{caseII}) turns to 
\begin{equation}
g_{\underset{s=2}{}}(r)=-m-\Lambda r^{2}+\frac{J^{2}}{4r^{2}}.
\label{caseIIb}
\end{equation}
%
%%%%%%%%%%%%%%%%%%%%%%%%%%%%%%%%%%%%%%%%%%%%%%%%%%%%%%%%%%%%%%%
\begin{figure}[tbh]
\centering
\includegraphics[width=0.4\linewidth]{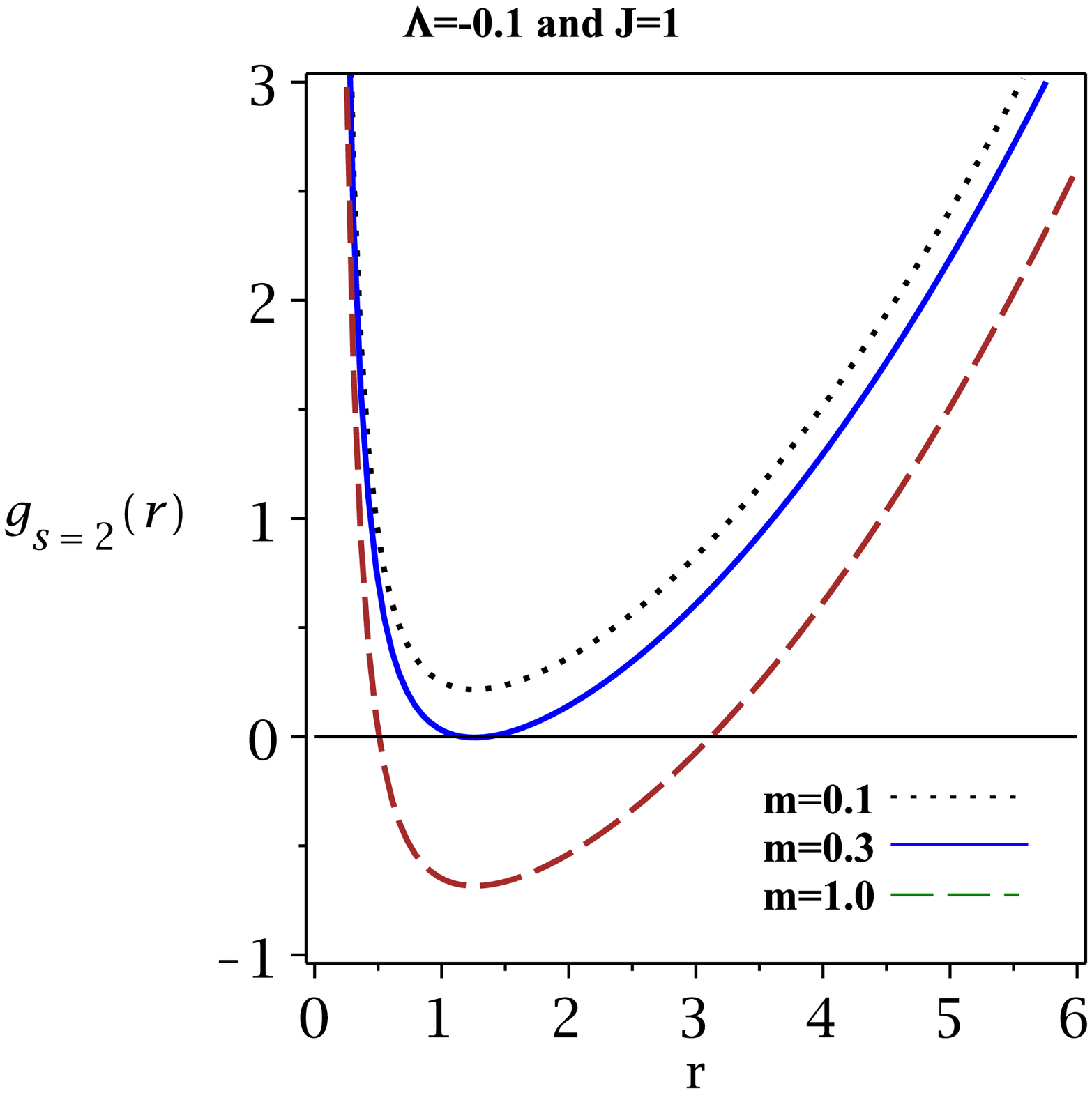} \includegraphics[width=0.4%
\linewidth]{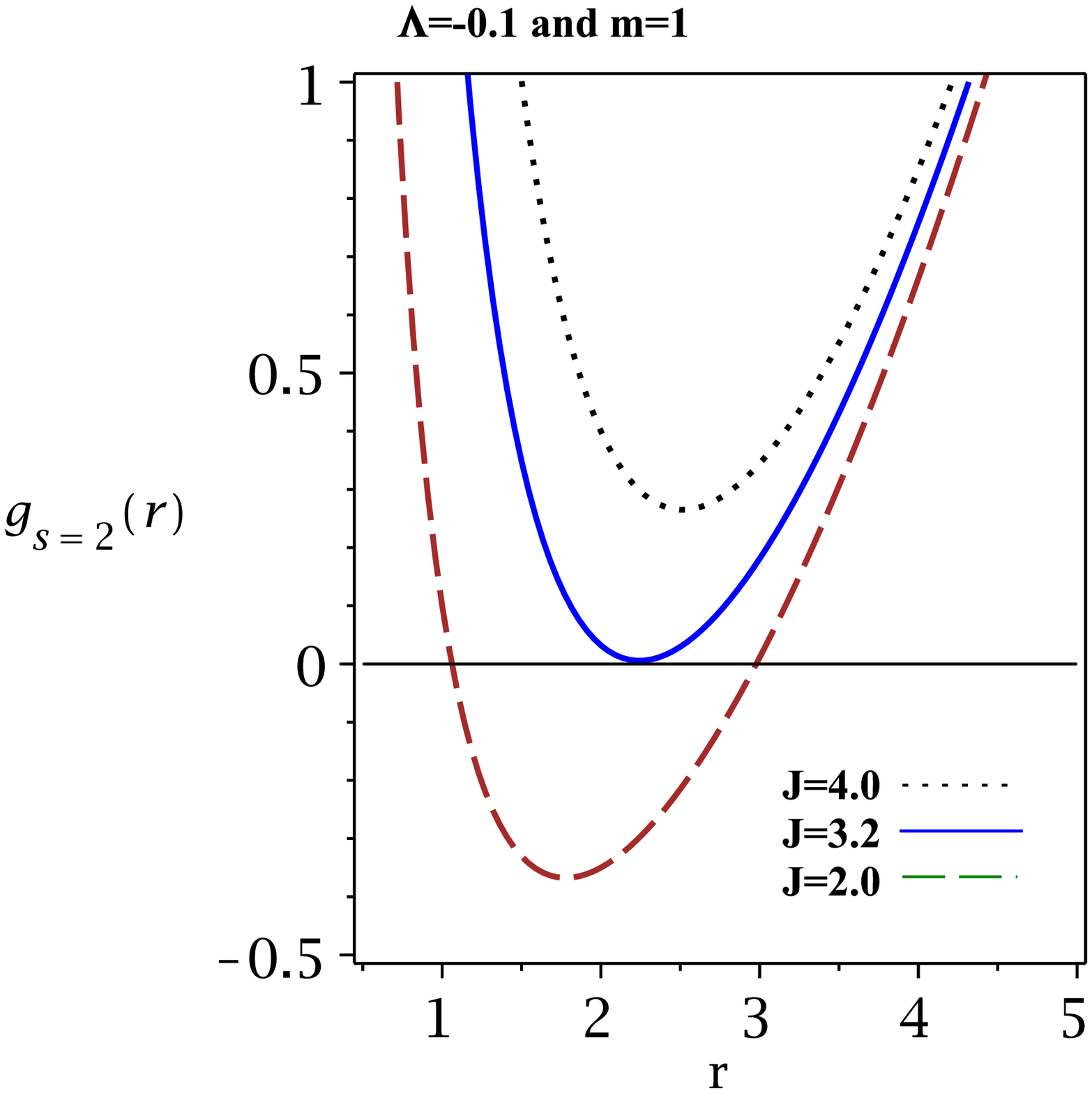}
\caption{$g_{\protect\underset{s=2}{}}(r)$ versus $r$ for different values
of $m$ (left panel) and $J=1$ (right panel).}
\label{Fig8}
\end{figure}
%%%%%%%%%%%%%%%%%%%%%%%%%%%%%%%%%%%%%%%%%%%%%%%%%%%%%%%%%%%%%%%

The Ricci and the Kretschmann scalars are given 
\begin{equation}
R_{0}=8\Lambda ,~~~~~\&~~~~~\mathcal{K}=4\Lambda ^{2},
\end{equation}%
and also $tt$-component of the Ricci tensor is $R_{tt}=\frac{16\Lambda r^{2}%
}{-4mr^{2}+J^{2}-4\Lambda r^{4}}$, in which indicates that there is a
curvature singularity at $r=0$, because the Ricci tensor diverges at $r=0$
(i.e., $\underset{r\rightarrow 0}{\lim }R_{tt}\rightarrow \infty $).

Fig. \ref{Fig8}, shows that by increasing (decreasing) the parameter $m$ ($J$%
), the black holes may have two horizons. In other words, massive $2D $
rotating-like AdS-black holes (or $2D$ slowly rotating-like AdS-black holes)
may have two horizons.

The Hawking temperature, mass, entropy, and also the heat capacity of the $%
2D $ rotating-like black holes are obtained as 
\begin{eqnarray}
T_{_{s=2}} &=&-\frac{\left( J^{2}+4\Lambda r_{+}^{4}\right) \left( \frac{%
r_{+}}{r_{0}}\right) }{8\pi r_{+}^{3}},~~~~~\&~~~~\ M_{_{s=2}}=\frac{%
J^{2}-4\Lambda r_{+}^{4}}{4r_{+}^{2}},  \notag \\
&& \\
S_{_{s=2}} &=&4\pi r_{0}\ln \left( \frac{r_{+}}{r_{0}}\right)
,~~~~~\&~~~~~C_{_{s=2}}=\frac{2\pi r_{0}\left( J^{2}+4\Lambda
r_{+}^{4}\right) }{J^{2}-4\Lambda r_{+}^{4}}.  \notag
\end{eqnarray}

\begin{figure}[tbh]
\centering
\includegraphics[width=0.4\linewidth]{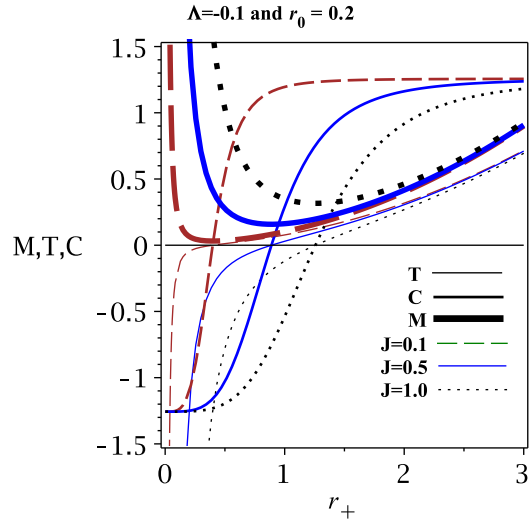}
\caption{$T$ (thin lines), $C$ (thick lines) and $M$ (thicker lines) versus $%
r_{+}$ for different values of $J$.}
\label{Fig9}
\end{figure}

The behavior of temperature, mass, and entropy in Fig. \ref{Fig9}, indicate
that $2D$ rotating-like AdS-black holes can be physical and enjoy thermal
stability in the range $r_{+}>r_{\text{root}}$. Another interesting result
is related to the rotating parameter effect ($J$). According to the obtained
results in Fig. \ref{Fig9}, the $2D$ rapidly rotating-like black holes are
physical and enjoy thermal stability in the large radius compared with the
slow case.

%%%%%%%%%%%%%%%%%%%%%%%%%%%%%%%%%%%%%%%%%%%%%%%%%%%%%%%%%%%%%%%
%%%%%%%%%%%%%%%%%%%%%%%%%%%%%%%%%%%%%%%%%%%%%%%%%%%%%%%%%%%%%%%

\section{Conclusions}

By considering the $2D$ Lifshitz-like spacetime, the black hole solutions
were extracted in $F(R)$ gravity. Our analysis indicated that the $2D$
Lifshitz-like AdS black holes with large radii could exist in this theory of
gravity.

Next, the thermodynamic quantities such as temperature, mass, and entropy
were obtained. The temperature and mass of the $2D$ Lifshitz-like AdS black
holes with large radii were positive simultaneously. In other words, the $2D$
Lifshitz-like large AdS black holes were physical. The heat capacity was
evaluated for the $2D$ Lifshitz-like black holes in $F(R)$ gravity. The
results indicated two critical points: i) the existence of one root for the
heat capacity. In this case, the small black holes were non-physical and
unstable. But the $2D$ Lifshitz-like black holes with large radii were
physical and enjoyed thermal stability. ii) although there was a divergent
point for the heat capacity. But it could not be real since the cosmological
constant has to be negative. Our results indicated a phase transition for
the $2D$ Lifshitz-like black holes between unstable non-physical and stable
physical cases. Briefly, according to the behavior of temperature and heat
capacity, the large AdS black holes in the $2D$ Lifshitz-like spacetime were
physical and enjoyed thermal stability.

In addition, the $2D$ Lifshitz-like black holes in $F(R)$ gravity might be
recovered two other interesting black holes by adjusting the Lifshitz-like
parameter. In other words, there was a correspondence between the $2D$
Lifshitz-like black holes in $F(R)$ gravity with Schwarzschild and or
rotating-like AdS black holes. These black hole solutions are:

I) \textbf{the $2D$ Schwarzschild AdS black holes}: the $2D$ Lifshitz-like
AdS black holes were reduced to the $2D$ Schwarzschild black holes when the
Lifshitz-like parameter was zero ($s=0$). For these black holes, the
thermodynamic quantities and heat capacity are obtained. The results
revealed that massive $2D$ Schwarzschild black holes might encounter two
horizons.

II) \textbf{the $2D$ rotating-like AdS black holes}: these solutions are
extracted by adjusting $s=2$. There were two horizons (inner and outer
horizons) for massive or slowly rotating-like AdS black holes in $2D$
Lifshitz-like spacetime. 
\begin{acknowledgements}
	
The author thanks University of Mazandaran.
		
\end{acknowledgements}

\section*{Data Availability}
Data sharing is not applicable to this article as no new data were created or analyzed in this
study.


\begin{thebibliography}{99}
\bibitem{expI} A. G. Riess, et al., Astron. J. \textbf{116}, 1009 (1998).

\bibitem{expIII} S. Perlmutter, et al., Astrophys. J. \textbf{517}, 565
(1999).

\bibitem{F(R)I} H. A. Buchdahl, Mon. Not. Roy. Astron. Soc. \textbf{150}, 1
(1970).

\bibitem{F(R)II} S. M. Carroll, V. Duvvuri, M. Trodden, and M. S. Turner,
Phys. Rev. D \textbf{70}, 043528 (2004).

\bibitem{F(R)III} S. Nojiri, and S. D. Odintsov, Phys. Rept. \textbf{505},
59 (2011).

\bibitem{Mod1} A. A. Starobinsky, Phys. Lett. B \textbf{91}, 99 (1980).

\bibitem{Mod2} I. Sawicki, and W. Hu, Phys. Rev. D \textbf{75}, 127502
(2007).

\bibitem{Mod4} L. Amendola, and S. Tsujikawa, Phys. Lett. B \textbf{660},
125 (2008).

\bibitem{Mod5} S. Tsujikawa, Phys. Rev. D \textbf{77}, 023507 (2008).

\bibitem{Mod5b} G. Cognola, E. Elizalde, S. Nojiri, S. D. Odintsov, L.
Sebastiani, and S. Zerbini, Phys. Rev. D \textbf{77}, 046009 (2008).

\bibitem{Mod6} S. Capozziello, E. Piedipalumbo, C. Rubano, and P.
Scudellaro, Astron. Astrophys. \textbf{505}, 21 (2009).

\bibitem{Mod9} A. V. Astashenok, S. Capozziello, and S. D. Odintsov, JCAP 
\textbf{12}, 040 (2013).

\bibitem{CapozzielloI} S. Capozziello, and A. Troisi, Phys. Rev. D \textbf{72%
}, 044022 (2005).

\bibitem{CapozzielloII} S. Capozziello, A. Stabile, and A. Troisi, Phys.
Rev. D \textbf{76}, 104019 (2007).

\bibitem{SoIII} S. Nojiri, and S. D. Odintsov, Phys. Rev. D \textbf{74},
086005 (2006).

\bibitem{SoV} K. Bamba, and S. D. Odintsov, JCAP \textbf{08}, 045 (2008).

\bibitem{SoVIII} S. H. Hendi, B. Eslam Panah, and S. M. Mousavi, Gen.
Relativ. Gravit. \textbf{44}, 835 (2012).

\bibitem{SoIX} S. G. Ghosh, S. D. Maharaj, and U. Papnoi, Eur. Phys. J. C 
\textbf{73}, 2473 (2013).

\bibitem{SoX} M. E. Rodrigues, E. L. B. Junior, G. T. Marques, and V. T.
Zanchin, Phys. Rev. D \textbf{94}, 024062 (2016).

\bibitem{SoXI} A. K. Mishra, M. Rahman, and S. Sarkar, Class. Quantum
Gravit. \textbf{35}, 145011 (2018).

\bibitem{SoXII} G. G. L. Nashed, and S. Capozziello, Phys. Rev. D \textbf{99}%
, 104018 (2019).

\bibitem{HoravaI} P. Horava, Phys. Rev. Lett. \textbf{102}, 161301 (2009).

\bibitem{HoravaII} P. Horava, Phys. Rev. D \textbf{79}, 084008 (2009).

\bibitem{FRHLI} J. Kluson, JHEP \textbf{11}, 078 (2009).

\bibitem{FRHLII} M. Chaichian, S. Nojiri, S. D. Odintsov, M. Oksanen, and A.
Tureanu, Class. Quantum Gravit. \textbf{27}, 185021 (2010).

\bibitem{FRHLIII} S. Carloni, M. Chaichian, S. Nojiri, S. D. Odintsov, M.
Oksanen, and A. Tureanu, Phys. Rev. D \textbf{82}, 065020 (2010).

\bibitem{FRHLIV} E. Elizalde, S. Nojiri, S. D. Odintsov, and D. S\'{a}ez-G%
\'{o}mez, Eur. Phys. J. C \textbf{70}, 351 (2010).

\bibitem{FRHLVII} A. J. L\'{o}pez-Revelles, R. Myrzakulov, and D. S\'{a}ez-G%
\'{o}mez, Phys. Rev. D \textbf{85}, 103521 (2012).

\bibitem{Kluson} J. Kluson, Phys. Rev. D \textbf{95}, 084026 (2017).

\bibitem{2II} F. David, Nucl. Phys. B \textbf{257}, 543 (1985).

\bibitem{2III} M. R. Douglas, and S. H. Shenker, Nucl. Phys. B \textbf{335},
635 (1990).

\bibitem{2IV} D. J. Gross, and A. A. Migdal, Phys. Rev. Lett. \textbf{64},
127 (1990).

\bibitem{2V} E. Brezin, and V. A. Kazakov, Phys. Lett. B \textbf{236}, 144
(1990).

\bibitem{2VI} V. A. Kazakov, I. K. Kostov, and A. A. Migdal, Phys. Rev.
Lett. \textbf{66}, 2051 (1991).

\bibitem{twoStI} E. Marinari, and G. Parisi, Phys. Lett. B \textbf{247}, 537
(1990).

\bibitem{twoStII} S. Nojiri, Phys. Lett. B \textbf{252}, 561 (1990).

\bibitem{2D1} M. Rotondo, and S. Nojiri, Mod. Phys. Lett. A \textbf{32},
1750149 (2017).

\bibitem{2D3} C. T. M. Ho, D. Su, R. B. Mann, and T. C. Ralph, Phys. Rev. D 
\textbf{94}, 081502(R) (2016).

\bibitem{2D4} D. Grumiller, M. Leston, and D. Vassilevich, Phys. Rev. D 
\textbf{89}, 044001 (2014).

\bibitem{2D7} K. Hotta, T. Kubota, and T. Nishinaka, Nucl. Phys. B \textbf{%
838}, 358 (2010).

\bibitem{2D9} D. Grumiller, and R. Meyer, Class. Quantum Gravit. \textbf{23}%
, 6435 (2006).

\bibitem{2D12} D. Grumiller, W. Kummer, and D. V. Vassilevich, Phys. Rept. 
\textbf{369}, 327 (2002).

\bibitem{2D13} S. Cacciatori, A. H. Chamseddine, D. Klemm, L. Martucci, W.
A. Sabra, and D. Zanon, Class. Quantum Gravit. \textbf{19}, 4029 (2002).

\bibitem{2D15} A. Kumar, and K. Ray, Phys. Rev. D \textbf{51}, 5954 (1995).

\bibitem{2D18} H. Kawai, and R. Nakayama, Phys. Lett. B \textbf{306}, 224
(1993).

\bibitem{2D19} S. Nojiri, and I. Oda, Nucl. Phys. B \textbf{406}, 499 (1993).

\bibitem{2D20} J. Russo, and A. A. Tseytlin, Nucl. Phys. B \textbf{382}, 259
(1992).

\bibitem{Lif4D} S. H. Hendi, B. Eslam Panah, and C. Corda, Can. J. Phys. 
\textbf{91}, 1 (2013).

\bibitem{HendiR} S. H. Hendi, R. Ramezani-Arani, and E. Rahimi, Phys. Lett.
B \textbf{805}, 135436 (2020).

\bibitem{NojiriOI} S. Nojiri, and S. D. Odintsov, EPL. \textbf{130}, 10004
(2020).

\bibitem{Calza} M. Calza, M. Rinaldi, and L. Sebastiani, Eur. Phys. J. C 
\textbf{78}, 178 (2018).

\bibitem{Bertipagani2021} M. Bertipagani, M. Rinaldi, L. Sebastiani, and S.
Zerbini, Phys. Dark Universe. \textbf{33}, 100853 (2021).

\bibitem{AMDI} A. Ashtekar, and A. Magnon, Class. Quantum Gravit. \textbf{1}%
, L39 (1984).

\bibitem{AMDII} A. Ashtekar, and S. Das, Class. Quantum Gravit. \textbf{17},
L17 (2000).

\bibitem{AMDIII} D. P. Jatkar, G. Kofinas, O. Miskovic, and R. Olea, Phys.
Rev. D \textbf{89}, 124010 (2014).

\bibitem{AMDIV} R. Emparan, and H. S. Reall, Living Rev. Relativ. \textbf{11}%
, 6 (2008).

\bibitem{Entropy} G. Cognola, E. Elizalde, S. Nojiri, S. D. Odintsov and S.
Zerbini, JCAP \textbf{02}, 25 (2005).

\bibitem{HeatI} B. Eslam Panah, Phys. Lett. B \textbf{787}, 45 (2018).

\bibitem{Achucarro} A. Ach\'{u}carro, and M. Ortiz, Phys. Rev. D \textbf{48}%
, 3600 (1993).
\end{thebibliography}
\end{document}